\documentclass[twocolumn,showpacs,preprintnumbers,amsmath,amssymb]{revtex4}

\usepackage{graphicx}
\usepackage{dcolumn}
\usepackage{bm}

\begin{document}

\preprint{APS/123-QED}

\title{Configuration mixing in the neutron-deficient $^{186-196}$Pb isotopes}

\author{V. Hellemans$^{1}$}
\altaffiliation{Present address : Universit\'e Libre de Bruxelles, Service de Physique Nucl\'eaire Th\'eorique, B-1050 Bruxelles, Belgium.}
\author{S. De Baerdemacker$^{1,2}$}%
\author{K. Heyde$^{1}$}
\affiliation{$^{1}$ Ghent University, Department of Subatomic and Radiation Physics, B-9000 Gent, Belgium\\
$^{2}$ University of Toronto, Department of Physics, Ontario M5S 1A7, Canada}

\date{\today}%

\begin{abstract}
In this article we report the results of detailed interacting boson model calculations with configuration mixing for the neutron-deficient Pb isotopes. Calculated energy levels and $B(E2)$ values for $^{188-196}$Pb are discussed and some care is suggested concerning the current classification on the basis of level systematics of the $4_1^+$ and $6_1^+$ states in $^{190-194}$Pb. Furthermore, quadrupole deformations are extracted for $^{186-196}$Pb and the mixing between the different families (0p-0h, 2p-2h, and 4p-4h) is discussed in detail. Finally, the experimental and the theoretical level systematics are compared.
\end{abstract}

\pacs{21.60.FW, 21.60.Ev}
\maketitle

\section{Introduction}
 The neutron-deficient Pb isotopes provide a unique region to unravel how the interactions between a large number of valence neutron holes in the $N$=82-126 shell and the multiparticle multihole proton excitations across the $Z$=82 closed shell give rise to a multitude of nuclear phenomena \cite{vanduppen84,wood92,andreyev00,julin01}. Whereas shell-model excitations dominate the low-energy spectrum near the doubly closed-shell nucleus $^{208}$Pb, collective modes of motion govern the properties of the nuclei in the vicinity of the neutron midshell. From a microscopic point of view, the experimentally observed dramatic lowering of two collective bands with decreasing neutron number is associated with the occurrence of 2p-2h and 4p-4h excitations across the closed $Z$=82 proton shell \cite{coster00}. These $n$p-$n$h excitations descend in energy mainly due to the pairing and the proton-neutron ($p$-$n$) quadrupole interaction \cite{heyde87}. Near the neutron midshell nucleus $^{186}$Pb ($N=104$), this effect is most pronounced due to the large number of valence neutron holes such that the states resulting from the 0p-0h, the 2p-2h, and the 4p-4h excitations occur within a small energy range. Hence, considerable mixing effects between the states belonging to different $n$p-$n$h excitation families may result. In the vicinity of $^{186}$Pb, the effects of mixing are largest and they gradually decrease when moving away from the neutron midshell. The efforts of several research groups have made detailed experimental data such as energy spectra, electromagnetic decay properties and isotopic shifts available (see Refs. \cite{dewitte07,pakarinen07,bujor07,grahn06,dracoulis05,dracoulis04,dewald03} for the most recent articles and Ref. \cite{julin01} and references therein for a review article) . The latter two observables are highly sensitive to the admixture of the different families of $n$p-$n$h excitations and provide reliable tools to test the validity of theoretical approaches in addition to the description of the energy spectrum.
 
 Several theoretical approaches are at one's disposal to study the Pb region. Early mean-field calculations understood the energy spectra of the Pb isotopes as resulting from the presence of spherical, prolate, and oblate minima in the energy surface \cite{may77,bengsston89,nazarewicz93,tajima93}.  Nowadays, beyond mean-field calculations which incorporate configuration mixing of angular momentum projected and particle-number projected self-consistent mean-field states provide a good description of various properties of the Pb isotopes \cite{bender04,rodriguez04}.

 A complementary approach is provided by an extended form of the interacting boson model (IBM) \cite{iachello87,frank94}. Whereas a shell-model calculation incorporating all necessary correlations for the neutron-deficient Pb isotopes is currently out of computational reach, the model space with multiparticle multihole excitations becomes tractable within the IBM truncation \cite{duval81,duval82}. The present article provides detailed results of calculations within the framework of the IBM with configuration mixing for the $^{186-196}$Pb isotopes, including results for the calculated energy levels, $B(E2)$ values and quadrupole deformations. These results complement previously published calculations on the midshell nucleus $^{186}$Pb \cite{pakarinen07} and on the calculated isotopic shifts in the neutron-deficient Pb isotopes \cite{dewitte07}. The present results for $^{188}$Pb in particular provide an update and extension of the results in Ref. \cite{hellemans05}.
\section{The formalism}
The IBM with configuration mixing allows the simultaneous treatment and mixing of several boson configurations that correspond to different particle-hole (p--h) shell-model excitations~\cite{duval81,duval82}. On the basis of intruder spin symmetry \cite{heyde92,coster96}, no distinction is made between particle and hole bosons. Hence, the shell-model space that includes 0p-0h, 2p-2h, and 4p-4h shell-model excitations corresponds to a $[N]\oplus[N+2]\oplus[N+4]$ boson space. Consequently, the Hamiltonian for three configuration mixing can be written as
\begin{align}
\hat{H}=&
\hat{P}^{\dag}_{N}\hat{H}^N_{\rm cqf}\hat{P}_{N}+
\hat{P}^{\dag}_{N+2}\left(\hat{H}^{N+2}_{\rm cqf}+\Delta^{N+2}\right)\hat{P}_{N+2}\nonumber\\
&+\hat{P}^{\dag}_{N+4}\left(\hat{H}^{N+4}_{\rm cqf}+\Delta^{N+4}\right)\hat{P}_{N+4}+
\hat{V}_{\rm mix}^{N,N+2}\nonumber\\
&+\hat{V}_{\rm mix}^{N+2,N+4}~,\label{eq:ibmhamiltonian}
\end{align}
where $\hat{P}_{N}$, $\hat{P}_{N+2}$, and $\hat{P}_{N+4}$ are projection operators
onto the $[N]$, the $[N+2]$, and the $[N+4]$ boson spaces, respectively, and
\begin{equation}
\hat{H}^i_{\rm cqf}=\varepsilon_i \hat{n}_d+\kappa_i \hat{Q}(\chi_i)\cdot\hat{Q}(\chi_i),\label{eq:cqfhamiltonian}
\end{equation}
is the consistent-$Q$ Hamiltonian \cite{warner83} with $i=N,N+2,N+4$, $\hat{n}_d$ the $d$ boson number operator, and
\begin{equation}
\hat{Q}_\mu(\chi_i)=[s^\dag\times\tilde{d}+ d^\dag\times s]^{(2)}_\mu+\chi_i[d^\dag\times\tilde{d}]^{(2)}_\mu~,\label{eq:quadrupoleop}
\end{equation}
the quadrupole operator. The parameters $\Delta^{N+2}$ and $\Delta^{N+4}$ can be associated with the energies needed to excite, respectively, two or four particles across a shell gap, corrected for the pairing interaction and a monopole effect~\cite{heyde87}. In general, the value of $\Delta^{N+4}$ is taken $2\Delta^{N+2}$ \cite{heyde91}. The operators $\hat{V}_{\rm mix}^{N,N+2}$ and $\hat{V}_{\rm mix}^{N+2,N+4}$ describe the mixing between, respectively, the $N$ and the $N+2$ configuration and the $N+2$ and $N+4$ configuration. They are defined as
\begin{equation}
\hat{V}_{\rm mix}^{i,i+2}=w_0^{i,i+2}[s^\dag\times s^\dag + s\times s]+w_2^{i,i+2} [d^\dag\times d^\dag+\tilde{d}\times \tilde{d}]^{(0)}\label{eq:vmix}~,
\end{equation}
where $i=N,N+2$. Conventionally, the Hamiltonian matrix is calculated in the $U(5)$ basis for diagonalization. This is the basis associated with the vibrational symmetry limit of the interacting boson model \cite{iachello87}. Hence one obtains the eigenvectors in the $[N]\oplus[N+2]\oplus[N+4]$ model space. A rotation of the Hamiltonian matrix to an intermediate basis that diagonalizes $\hat{H}_{\rm cqf}^{N}$, $\hat{H}_{\rm cqf}^{N+2}$, and $\hat{H}_{\rm cqf}^{N+4}$ in the different subspaces $[N]$, $[N+2]$, and $[N+4]$ respectively, results in the unperturbed energies and the interaction matrix elements that couple these unperturbed states. In other words, in the intermediate basis we obtain the energy levels (and bands) in the 0p-0h, 2p-2h, and 4p-4h subspaces in the absence of configuration mixing. The matrix elements of $\hat{V}_{\rm mix}^{N,N+2}$ and $\hat{V}_{\rm mix}^{N+2,N+4}$ expressed in the intermediate basis give the mixing strengths between these unperturbed states. We refer the reader to Ref. \cite{hellemans05} for a more detailed discussion.

The $E2$ transition operator for three-configuration mixing is defined as
\begin{equation}
\hat{T}^{(\rm E2)}_\mu=\sum_{i=N,N+2,N+4} e_i \hat{P}_i^\dag\hat{Q}_\mu(\chi_i)\hat{P}_i~,\label{eq:e2operator}
\end{equation}
where the $e_i$ ($i=N,N+2,N+4$) are the effective boson charges (\ref{eq:quadrupoleop}).

\begin{table}[!tb]
\begin{center}
\caption{Values of the parameters $\Delta^{N+2}$ and $\Delta^{N+4}$ for $^{186-196}$Pb. }\label{tab:parameters1}
\begin{tabular}{lllllll}
\hline\hline $A$ & 186 & 188 & 190 & 192 & 194 & 196\\
\hline $\Delta^{N+2}$ (MeV) &  2.129 & 1.923 & 1.816 & 1.744 & 1.800 & 1.865\\
     $\Delta^{N+4}$ (MeV) &  4.258 & 3.846 & 3.632 & 3.488 & 3.600 & 3.730\\
\hline\hline
\end{tabular}
\caption{Parameters of the IBM Hamiltonian \eqref{eq:ibmhamiltonian} and the $E2$ operator \eqref{eq:e2operator} for $^{186-196}$Pb. The values for $\varepsilon_i$, $\kappa_i$, and $\chi_i$ were taken from Ref. \cite{fossion03}. }\label{tab:parameters2}
\begin{tabular}{llll}
\hline\hline            & $N$ & $N+2$ & $N+4$\\
\hline $\varepsilon_i$ (MeV) & 0.92    & 0.51    & 0.55\\
       $\kappa_i$	(MeV)   & 0       & -0.014  & -0.020\\
       $\chi_i$           & 0       & 0.515   & -0.680\\
$w^{i,i+2}_0$=$w^{i,i+2}_2$ (MeV) & 0.018   & 0.018 & \\
$e_i$ ($e$b)		& 0.110   & 0.140   & 0.170\\
\hline\hline	
\end{tabular}
\end{center}
\end{table}
%
Starting from the schematic fit performed by Fossion \textit{et al.} \cite{fossion03}, we slightly adjusted certain parameters in the Hamiltonian \eqref{eq:ibmhamiltonian}. The values of $\varepsilon_i$, $\kappa_i$, and $\chi_i$ ($i=N, N+2, N+4$), which were determined on the basis of intruder spin symmetry \cite{fossion03}, remain unaltered, except for $\varepsilon_N$ which was changed little for a better description of the regular $2^{+}$ state. The value of the mixing parameters $w_0^{N,N+2}$, $w_2^{N,N+2}$, $w_0^{N+2,N+4}$, and $w_2^{N+2,N+4}$ were refined to describe the recently measured $B(E2)$ values in $^{186,188}$Pb \cite{grahn06} as well as possible. These $B(E2)$ values provide a good tool to fix the mixing parameters because they are sensitive to the precise structure of the wave functions which in turn is sensitive to the mixing. All these  parameters are taken constant for $^{186-196}$Pb. For the parameter $\Delta^{N+2}$, we introduced a slight variation to obtain a better description of the slope of the energy levels throughout the isotopic chain. 
All parameters are summarized in Table \ref{tab:parameters1} and \ref{tab:parameters2}. This set of parameters was already used for the calculations on $^{186}$Pb \cite{pakarinen07} and for the calculation of the isotopic shifts \cite{dewitte07}.\\

In the subsequent section, we present the results of our calculations and confront them with the available experimental data. In $^{186,188}$Pb, recent experiments \cite{dracoulis04,grahn06,pakarinen07} have provided extensive data on excitation spectra and the electromagnetic properties, rendering a thorough test for theoretical approaches. For the heavier $^{190-196}$Pb isotopes, we compare the results of the calculations to the experimental level schemes. In general, the low-lying excited states for these isotopes were organized and interpreted on the basis of level systematics. It will be shown that the classification of these low-lying states should be handled with care.\\

In the following, the theoretical bands are defined by following the $E2$ decay starting from the high-spin states using the criterion that in-band $B(E2)$ values must be larger than the corresponding interband $B(E2)$ values.

\section{Results}

\begin{table}[!tb]
\begin{center}
\caption{Comparison between the experimental and the calculated $B(E2)$ ratios for the interband $J\rightarrow J$ and $J\rightarrow J-2$ transitions between Band II and Band I in $^{188}$Pb. The ratios are normalized to the corresponding in-band $J\rightarrow J-2$ transition in Band II, except for the ratios starting from $2_2^+$ which were normalized to the $2_2^+\rightarrow 0_1^+$ transition. The experimental ratios were extracted from the branching ratios reported in \cite{dracoulis04}.}\label{tab:ratio188}
\begin{tabular}{c c c c}
\hline\hline
$J^\pi_{initial}$&  $J^\pi_{final}$ & $B(E2)$ ratio (IBM) & $B(E2)$ ratio (exp.) \\
\hline\hline
$8_2^+$	 & $8_1^+$ & 31.8 & 47.4 \\
	& $6_1^+$ & 1.6 & 0.6 \\
$6_2^+$	 & $6_1^+$ & 40.0 & 27.1 \\
	& $4_1^+$ & 0.3 & 1.5 \\
$4_2^+$	 & $4_1^+$ & 96.0 & 87.9 \\
	& $2_1^+$ & 3.3 & 7.8 \\
$2_2^+$	 & $2_1^+$ & 1947.2 & 14742.6\\
	& $0_2^+$ & 642.3 & $<$533.9\\
	& $0_3^+$ & 2254.6 & $<$28088.6\\
\hline\hline 
\end{tabular}
\end{center}

\end{table}
%
\begin{figure}[!tbp]
\begin{center}
\includegraphics[angle=-90,width=\columnwidth]{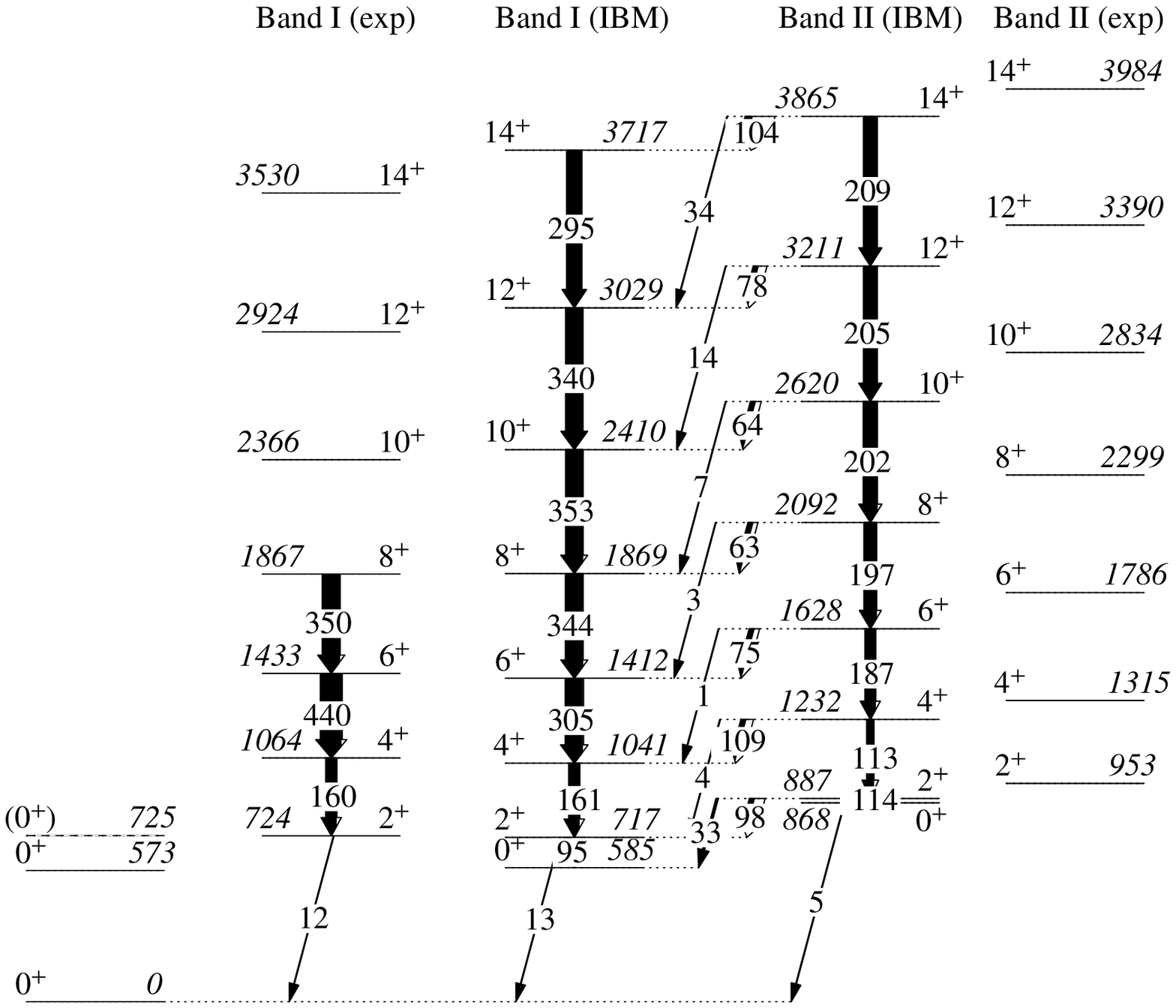}
\end{center}
\caption{Experimental level scheme and calculated energy levels for $^{188}$Pb. The arrows indicate the $B(E2)$ value of the transition and are expressed in Weisskopf units (W.u.). The widths are proportional to the $B(E2)$ value. Experimental data were taken from Refs. \cite{dracoulis04,grahn06}.}\label{fig:levelscheme188}
\end{figure}
%
For a detailed study of the calculated energy levels in $^{186}$Pb and discussion, we refer the reader to Ref. \cite{pakarinen07}. We briefly summarize here that the energy levels and the $B(E2)$ values, obtained in recoil distance Doppler-shift lifetime measurements \cite{grahn06}, are well described by our calculations. Moreover, the characteristic experimentally observed pattern of strong $J\rightarrow J$ interband $E2$ transition rates compared to relatively weak $J\rightarrow J-2$ interband $E2$ transition rates \cite{pakarinen07} is reproduced. Complementary results such as electric quadrupole moments and mixing strengths for $^{186}$Pb are included later in this discussion.

The experimental level scheme \cite{dracoulis04,grahn06} and the calculated energy levels for $^{188}$Pb are presented in Fig. \ref{fig:levelscheme188}. The calculated regular $2^+$ state, which is not shown, occurs at an excitation energy of 1061 keV. The experimental energies in Band I and II and the $B(E2)$ values, obtained from lifetime measurements \cite{grahn06}, are rather well reproduced by our calculations. Band I has a predominant 4p-4h character whereas Band II is mainly 2p-2h. The interband transitions from the theoretical Band II to Band I exhibit the same pattern as in $^{186}$Pb where the interband $J\rightarrow J$ $E2$ transition rates are considerably larger than the interband $J\rightarrow J-2$ $E2$ transition rates \cite{pakarinen07}. Table \ref{tab:ratio188} presents the experimental $B(E2)$ ratios extracted from the branching ratios given in Dracoulis \textit{et al.} \cite{dracoulis04} as well as the corresponding calculated values. From this table, we see that the calculated ratios reproduce the pattern of experimentally deduced ratios. The present results provide an update and extension to previous studies of this isotope \cite{fossion03,hellemans05} to meet the recent experimental results on the $B(E2)$ values \cite{grahn06}.

\begin{figure}[!tbp]
\begin{center}
\includegraphics[angle=-90,width=\columnwidth]{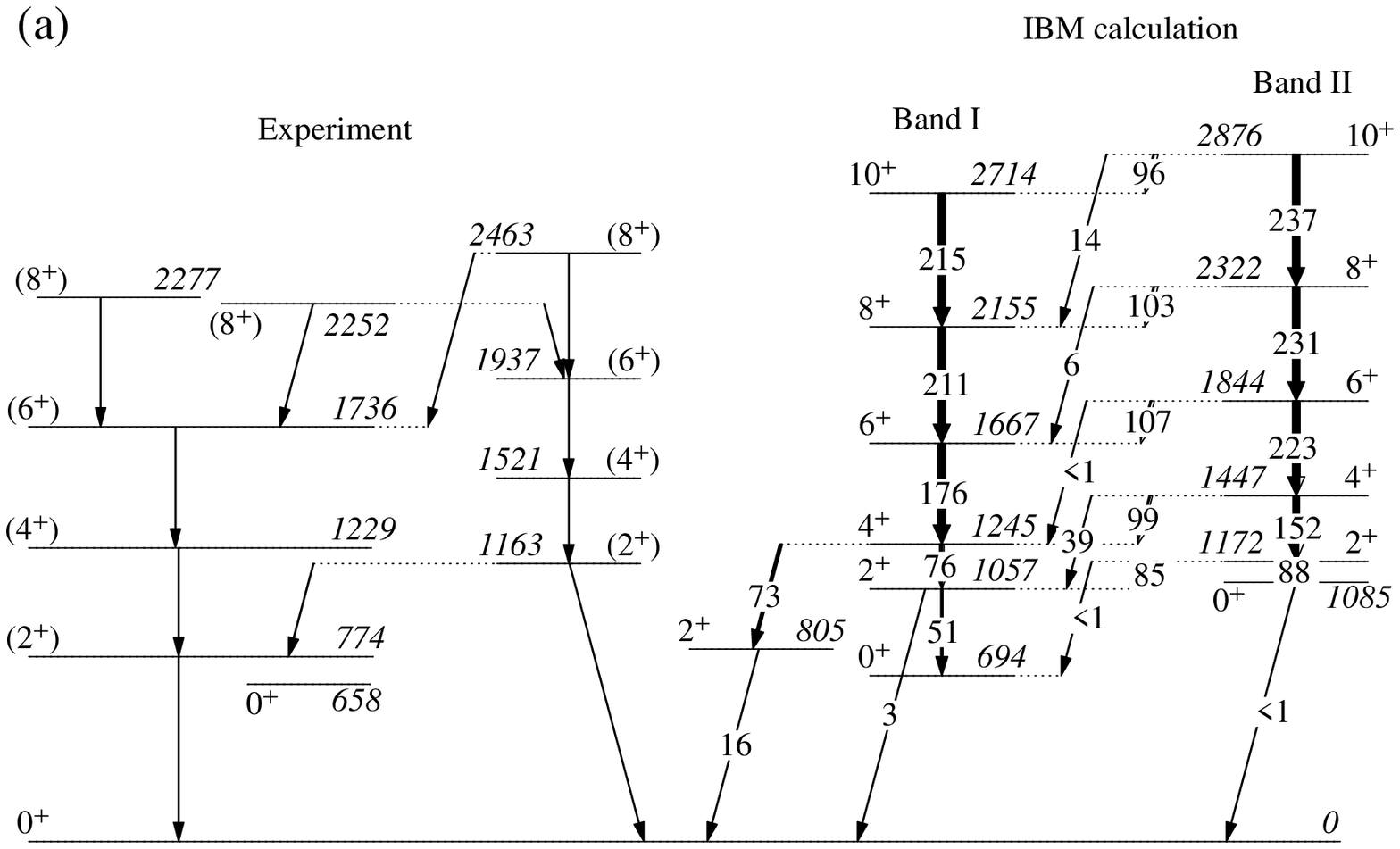}
\end{center}
\begin{center}
\includegraphics[angle=-90,width=\columnwidth]{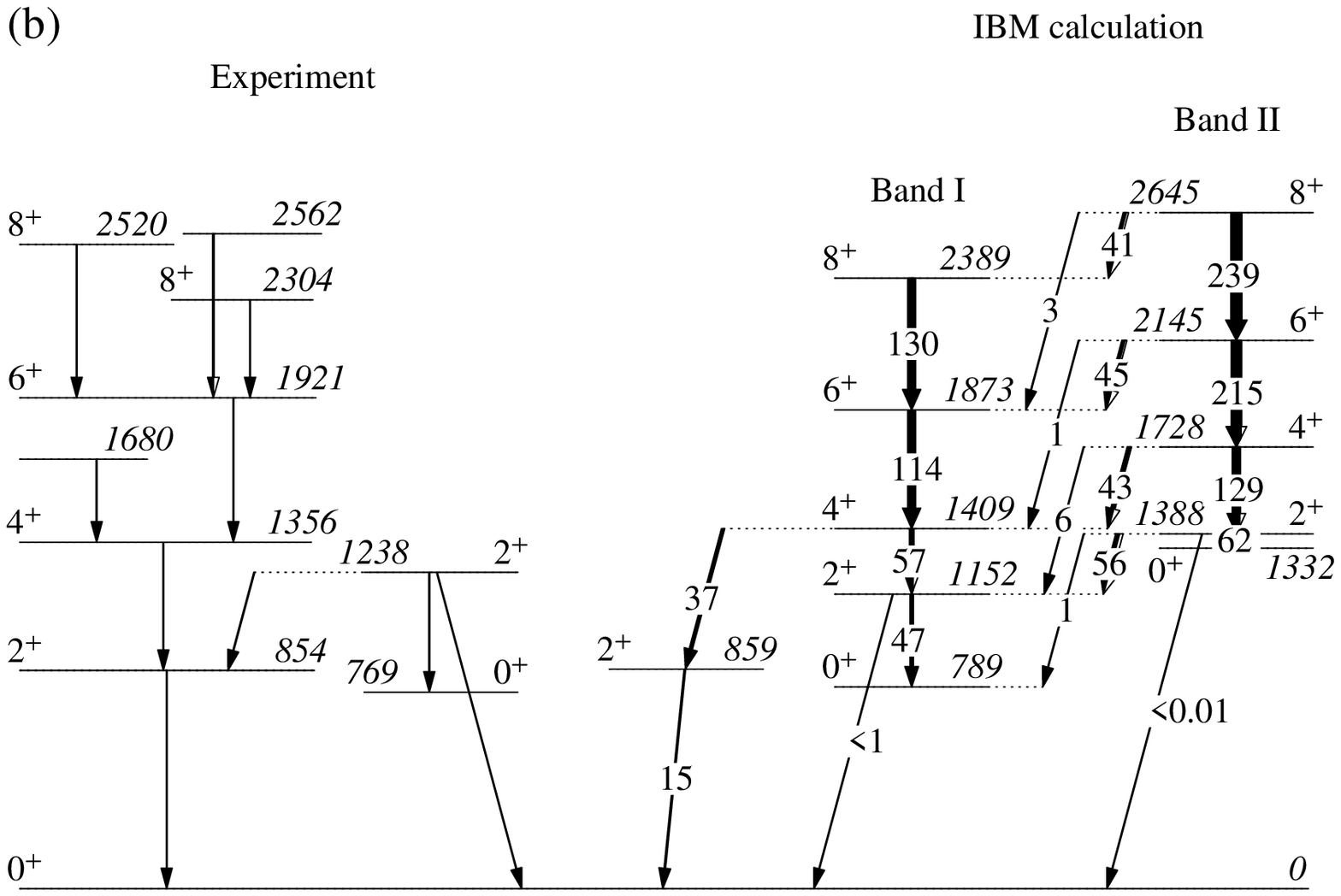}
\end{center}
\caption{Experimental and calculated energy scheme for $^{190}$Pb (a) and $^{192}$Pb (b). Experimental data for $^{190}$Pb were taken from Ref. \cite{dracoulis98} and data for $^{192}$Pb from Refs. \cite{vanduppen87,plompen93,mcnabb97}. The arrows in the experimental decay pattern indicate the connecting transitions. The calculated $B(E2)$ values are given in W.u. and the width of the arrows is proportional.}\label{fig:levelscheme190/2}
\end{figure}
\begin{figure}[!tbp]
\begin{center}
\includegraphics[angle=-90,width=\columnwidth]{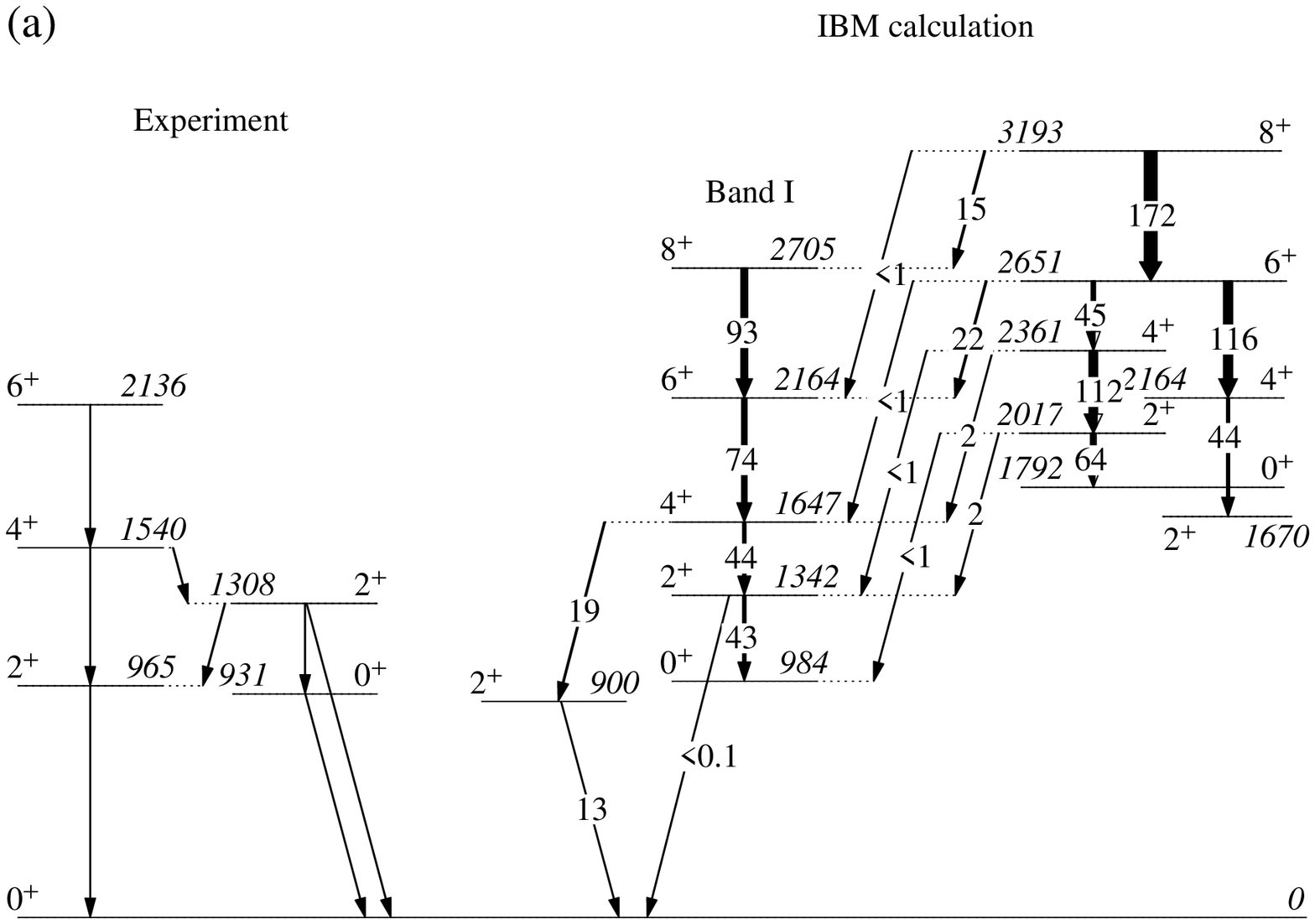}
\end{center}
\begin{center}
\includegraphics[angle=-90,width=\columnwidth]{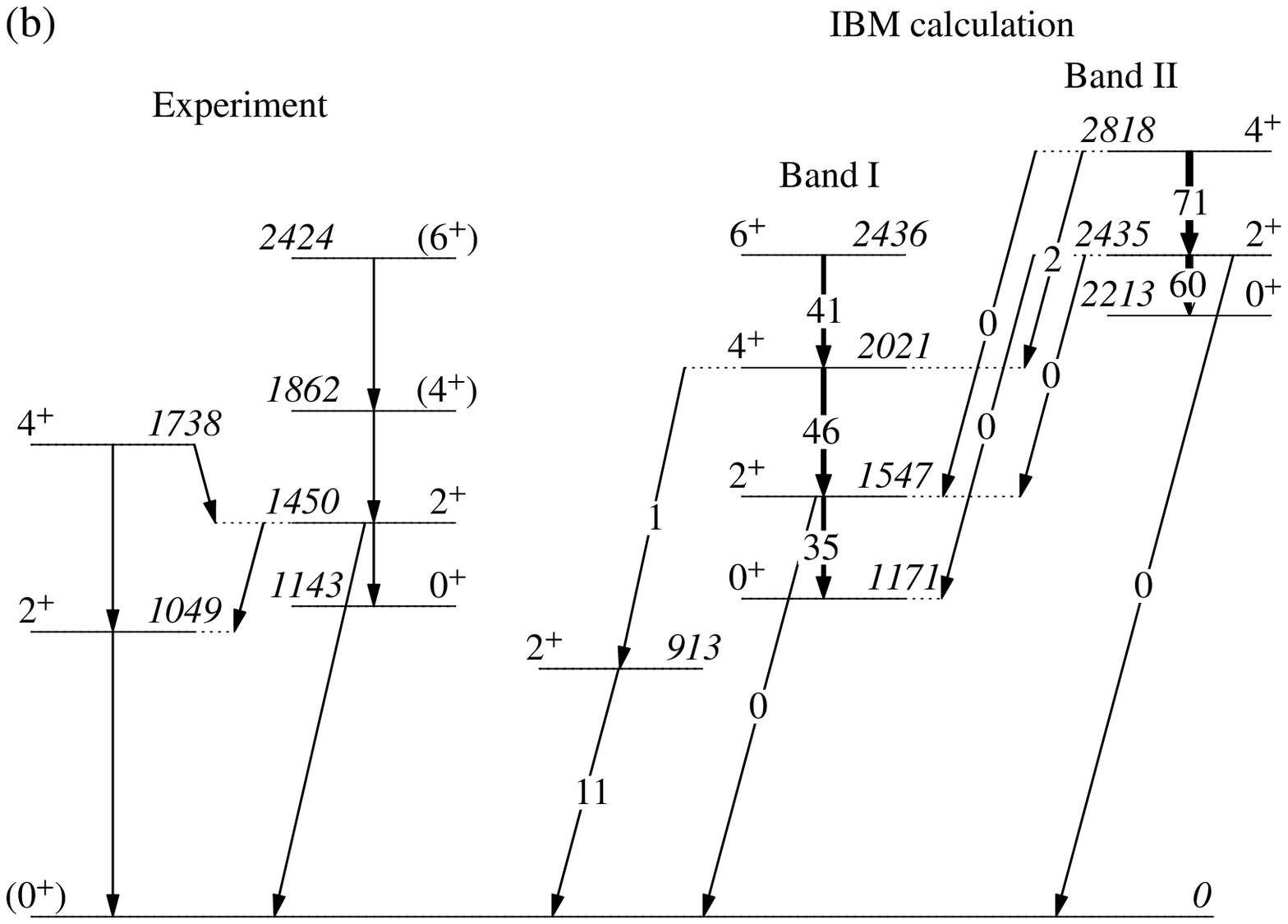}
\end{center}
\caption{Experimental and calculated energy scheme for $^{194}$Pb (a) and $^{196}$Pb (b). Experimental data for $^{194}$Pb were taken from Refs. \cite{vanduppen87,dracoulis05, kaci02} and data for $^{196}$Pb from Refs. \cite{vanduppen87,dracoulis05,penninga87}. The arrows in the experimental decay pattern indicate the connecting transitions. The calculated $B(E2)$ values are given in W.u. and the width of arrows is proportional.}\label{fig:levelscheme194/6}
\end{figure}

Figure \ref{fig:levelscheme190/2} and \ref{fig:levelscheme194/6} display the experimental and the calculated energy levels for $^{190}$Pb-$^{196}$Pb. For $^{190}$Pb [Fig. \ref{fig:levelscheme190/2}(a)] , the experimental $0^+_2$ state at 658 keV is interpreted as a 2p-2h intruder state \cite{dendooven89} and the side band built on the 1163 keV $2_2^+$ state is identified as a prolate band built on the 4p-4h proton excitation \cite{dracoulis98}. The experimental $8^+$ states are related to strongly mixed spherical, oblate, and prolate states. In general, the yrast $2^+_1$, $4^+_1$, and $6^+_1$ states were associated with spherical states on the basis of level systematics \cite{dracoulis98,dracoulis99,julin01}.

In the IBM calculation, Band I is of mainly 2p-2h character and is strongly mixed with Band II which mainly has a 4p-4h character. The energies of Band II are in reasonable agreement with those of the experimental band assigned as a prolate band \cite{dracoulis98}. Likewise, the calculated $2_1^+$ and the $0^+_2$ state are in good agreement with the corresponding experimental levels. 

In our IBM calculation, the regular configuration is described with the Hamiltonian of the $U(5)$ symmetry limit. This implies that the regular $2^+$ IBM state is described by N-1 $s$ bosons and one $d$ boson ($J$=2). Higher-lying excited states result from angular momentum coupling of additional $d$ bosons and $s$ bosons. Because the regular excitation mode is understood as resulting from neutron pair breaking mainly, the IBM description for the regular states is only valid up to the $2^+$ state. Hence, it is important to verify to what extent the higher-lying unphysical $U(5)$ states mix with the 2p-2h and 4p-4h states. From Table  \ref{tab:appendix} in Appendix \ref{appendix}, which contains the weights of the wave functions in each configuration subspace, we notice that the $U(5)$ states with $J>2$ mix little with the states in Band I and Band II in $^{190}$Pb.

From inspection of Fig. \ref{fig:levelscheme190/2}(a), it becomes clear that the energies of the $4^+_1$ and the $6^+_1$ state in Band I lie very close to those of the corresponding yrast states determined experimentally. Moreover, our calculations indicate a strong $E2$ transition from the $4_1^+$ to the $2_1^+$ state. This might suggest that the observed $4^+_1$ and $6^+_1$ yrast states are related to 2p-2h intruder states. However, the IBM calculation lacks information on the regular states with $L>2$. If these experimental states are of spherical nature, our calculations suggest they will be strongly mixed with the states from the 2p-2h intruder configuration.

A similar situation arises in $^{192}$Pb \cite{vanduppen87,plompen93,mcnabb97} [Fig. \ref{fig:levelscheme190/2}(b)]. Experimentally, the 769 keV $0^+_2$ and 1238 keV $2^+_2$ state are identified as 2p-2h states whereas the yrast $2^+_1$, $4^+_1$, and $6^+_1$ are associated with regular states on the basis of level systematics \cite{dracoulis99,julin01}. In the IBM calculation, Band I has a dominant 2p-2h character and Band II a dominant 4p-4h character. Again, the energies of the calculated $4^+_1$ and $6^+_1$ states in Band I lie close to their experimental counterparts that were interpreted as regular states from level systematics. The calculated $E2$ decay exhibits a relatively strong $B(E2)$ value from the mainly 2p-2h $4^+_1$ in Band I to the regular $2^+_1$ and therefore a conclusion similar as for $^{190}$Pb can be drawn.
%
\begin{figure}[!tbp]
\begin{center}
\includegraphics[width=\columnwidth]{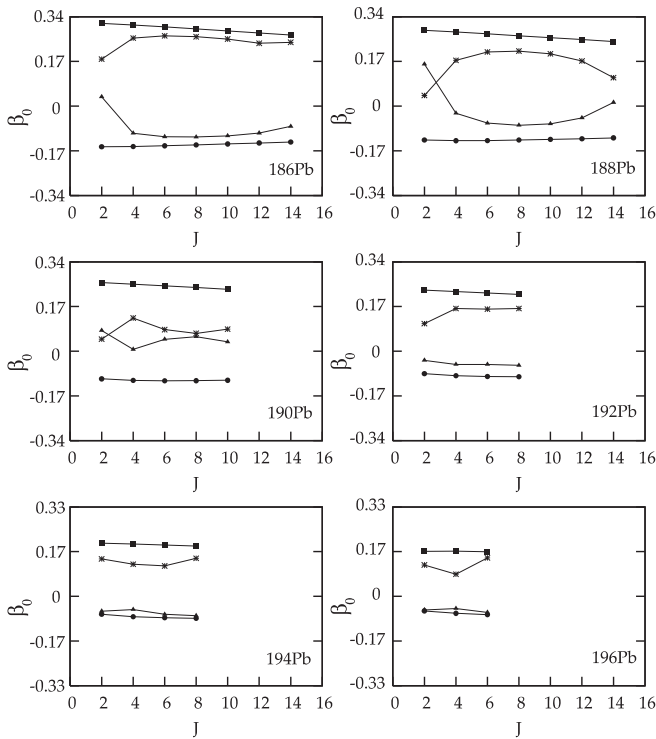}
\caption{Quadrupole deformations $\beta_0$ extracted from the spectroscopic quadrupole moments in the assumption of a $K=0$ band for $^{186-196}$Pb (\ref{eq:bzero}). Results for the 2p-2h unperturbed band are represented with $\bullet$, and results for the 4p-4h unperturbed band are represented with $\blacksquare$. The mixed (mainly) 2p-2h band and the mixed (mainly) 4p-4h band are represented by $\blacktriangle$ and $\ast$ respectively. The number of states shown diminishes with increasing neutron number because the model space becomes smaller and less-high spin states can be calculated in a reliable way. }\label{fig:multiplot_iqm}
\end{center}
\end{figure}

The theoretical picture gets more complicated in $^{194}$Pb [Fig. \ref{fig:levelscheme194/6}(a)]. The calculated Band I still has a dominant 2p-2h structure. On the contrary, the picture for the higher-lying calculated states displayed on the rightmost side in Fig. \ref{fig:levelscheme194/6}(a) is less clear. The low-spin states of the lowest band built on the 4p-4h proton excitation are very mixed up with the low-spin states of the second band built on the 2p-2h proton excitation, which is reflected in Table \ref{tab:appendix} (given in Appendix \ref{appendix}). As our main interest goes out to Band I, we do not display all connecting transitions starting from these higher-lying states (see Ref. \cite{hellemans07b} for all connecting transitions). The $4_1^+$ state in Band I is 22\% mixed with the $4^+$ state of the $U(5)$ limit, which was chosen to describe the regular configuration with $N$ bosons. This mixing only shifts the $4_1^+$ state by approximately 100 keV from its unperturbed energy but can affect the $B(E2; 4^+_1\rightarrow 2^+_1)$ transition. Nevertheless, the calculated energies of the $4^+_1$ and $6^+_1$ states again lie close to those observed experimentally \cite{dracoulis05,kaci02} and again a rather strong $E2$ transition (although probably somewhat overestimated, due to the mixing) from the theoretical $4^+_1$ to $2^+_1$ state  results. Hence the 2136 keV $6^+_1$ state and the 1540 keV $4^+_1$ state may be associated with the corresponding states of 2p-2h character in Band I.

The isotope $^{196}$Pb [Fig. \ref{fig:levelscheme194/6}(b)] is of key importance for our calculations. In this isotope, an even-spin side band exhibiting the characteristics of a collective band has been measured and is understood as a 2p-2h intruder band  \cite{vanduppen87,penninga87,dracoulis05}. Similar to the previously discussed isotopes, Band I is a mainly 2p-2h band and Band II is a mainly 4p-4h band. The calculated energies of Band I are in good agreement with the experimental energies of the even-spin side band. Moreover, the calculated $B(E2)$ value from the $4_1^+$ state in Band I to the regular $2_1^+$ state, which is strong in $^{190-194}$Pb, drops considerably in $^{196}$Pb. This makes our interpretation for the $A$=190-196 Pb isotopes consistent with the experimentally observed $E2$ decay pattern. It is important to mention that the $4_1^+$ state in $^{196}$Pb is mixed with the nonphysical $4^+$ state in the regular band (see Table \ref{tab:appendix} in Appendix \ref{appendix}). This implies that the $B(E2;4^+_1\rightarrow 2^+_1)$ transition is most likely overestimated.\\

We proceed with a discussion on the quadrupole moments for $^{186-196}$Pb. The spectroscopic quadrupole moment can be calculated as
\begin{align}
Q^{(s)}(J,k)&=\sqrt{\frac{16\pi}{5}}\langle J,M=J,k|\hat{T}^{(E2)}_{\mu=0}|J,M=J,k\rangle~,\nonumber\\
&=\sqrt{\frac{16\pi}{5}}\frac{\langle JJ20|JJ\rangle}{\sqrt{2J+1}}\langle J,k||\hat{T}^{(E2)}||J,k\rangle~,
\end{align}
where $(J,k)$ denotes the $k^\textrm{th}$ state with total angular momentum $J$.
It is convenient to relate this spectroscopic quadrupole moment to the qua\-dru\-pole moment in the intrinsic frame of the nucleus to extract the quadrupole deformation $\beta_0$ of the nucleus. The IBM, however, is formulated within the laboratory frame. Hence, we should compare the calculated spectroscopic quadrupole moment with the rigid rotor model to extract information in the intrinsic frame \cite{rowe70}. In the following, we assume the lowest 2p-2h and 4p-4h bands to be $K=0$ bands. Then, the intrinsic quadrupole moment $Q^{(i)}_{K=0}(J,k)$ and the quadrupole deformation $\beta_0$ can be extracted using the expressions
\begin{align}
&Q^{(i)}_{K=0}(J,k)=-\frac{(2J+3)}{J}Q^{(s)}(J,k)~,\label{eq:kzero}\\
&\beta_0=\frac{\sqrt{5\pi}}{3ZR_0^2}Q^{(i)}_{K=0}(J,k)\label{eq:bzero}~,
\end{align}
where $Z$ denotes the number of protons and $R_0=1.2A^{1/3}$fm is the nuclear radius. 
The results in the intrinsic frame are presented in Fig. \ref{fig:multiplot_iqm}. For the unperturbed 2p-2h and 4p-4h bands, we find an oblate and prolate deformation, respectively. This result is consistent with results of beyond mean-field calculations \cite{bender04,rodriguez04}. The magnitude of the deformation of both the 2p-2h and the 4p-4h band decreases with increasing neutron number. Within the unperturbed bands, $\beta_0$ stays approximately constant as a function of $J$ whereas large mixing effects can occur in the mixed bands. More specifically, the quadrupole deformations for the mixed bands cross at low $J$ in $^{188}$Pb and $^{190}$Pb. This can be understood from the definition of the theoretical bands (which were constructed by following the $E2$ decay) and the large mixing effects in the low $J$ states of these nuclei. For the other isotopes, mixing effects are less dominant and the deviation between the deformation of the mixed and unmixed bands becomes smaller.\\

\begin{figure}[!t]
\begin{center}
\includegraphics[width=\columnwidth]{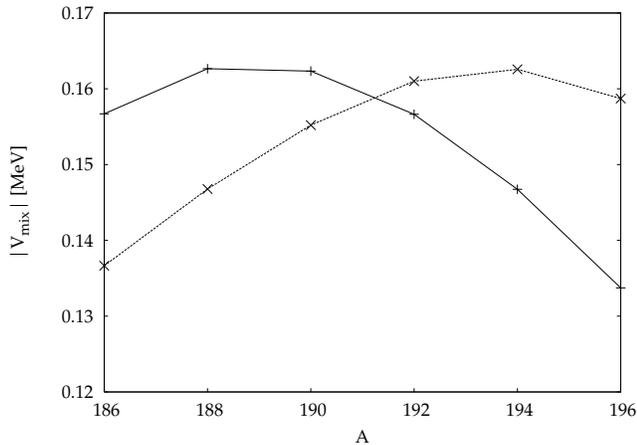}
\caption{Mixing strengths between the lowest unperturbed $0^+$ 0p-0h and 2p-2h states ($+$) and between the lowest unperturbed $0^+$ 2p-2h and 4p-4h states ($\times$) for $^{186-196}$Pb. }\label{fig:vmixzeros}
\end{center}
\end{figure}

Detailed information on the mixing can be obtained from the mixing strengths and from the $B(E2)$ values.
The knowledge of the intermediate bases (see Ref. \cite{hellemans05}) allows us to calculate the expectation value of $\hat{V}_{\rm mix}^{N,N+2}$ and $\hat{V}_{\rm mix}^{N+2,N+4}$. Expressed in the intermediate bases, they give the interaction strength between the unperturbed energy levels and can be compared with phenomenological band-mixing calculations \cite{dracoulis94,dracoulis98,allatt98,page01,dracoulis04} in a straightforward way.
The mixing strengths between the three lowest $0^+$ states in $^{186-196}$Pb can be found in Fig. \ref{fig:vmixzeros}.

Additional information on the mixing is obtained from the $B(E2)$ values. The knowledge of the intermediate bases allows us to relate the reduced matrix element of the $E2$ transition from an initial state $J_i(i)$ to a final state $J_f(f)^{1}$ to the reduced matrix elements of the corresponding $E2$ transitions in the unperturbed bands. 
\begin{widetext}
\begin{align}
\langle J_f,f||T^{(E2)}||J_i,i\rangle&=\sum_{k,p,s,s'}a_{ki}(J_i)a_{pf}(J_f)\tilde{b}^{\nu}_{sk}(J_i)\tilde{b}^{\nu}_{s'p}(J_f)~_\nu'\langle J_f,s'||T^{(E2)}||J_i,s\rangle'_\nu\Big|_{\nu=N}\nonumber\\
&\quad+\sum_{l,q,t,t'}a_{li}(J_i)a_{qf}(J_f)\tilde{b}^{\nu}_{tl}(J_i)\tilde{b}^{\nu}_{t'q}(J_f)~'_{\nu}\langle J_f,t'||T^{(E2)}||J_i,t\rangle'_{\nu}\Big|_{\nu=N+2}\nonumber\\
&\quad+\sum_{m,r,u,u'}a_{mi}(J_i)a_{rf}(J_f)\tilde{b}^{\nu}_{um}(J_i)\tilde{b}^{\nu}_{u'r}(J_f)~'_{\nu}\langle J_f,u'||T^{(E2)}||J_i,u\rangle'_{\nu}\Big|_{\nu=N+4}~,\label{eq:nr2}
\end{align}
where $\mathcal{A}$ is the matrix that diagonalizes the Hamiltonian \eqref{eq:ibmhamiltonian} in the $U(5)$ $[N]\oplus[N+2]\oplus[N+4]$ boson basis and $\mathcal{B^\nu}$ ($\nu=N, N+2, N+4$) are the three matrices that diagonalize the  respective $\hat{H}^\nu_{\rm cqf}$ \eqref{eq:cqfhamiltonian} in the corresponding $U(5)$ $[\nu]$ boson basis. From the latter diagonalization, we obtain the unperturbed energies and the corresponding eigenvectors. These eigenvectors span the intermediate basis $|J,r\rangle'_\nu$ with $J$ the total angular momentum and $r$ the rank number of the state in the unperturbed band.
Hence, we can filter out those reduced $E2$ matrix elements in the unperturbed bands that contribute more than 90\% to the reduced $E2$ matrix element of the $J_i(i)\rightarrow J_f(f)$ transition. For the in-band transitions in $^{186-192}$Pb, it turns out that the main contribution always consists of the terms with the corresponding transitions in the lowest unperturbed 2p-2h band and in the lowest unperturbed 4p-4h band. So, we may write
\begin{align}
\langle J_f,f||T^{(E2)}||J_i,i\rangle &\cong \sum_{l,q} a_{li}(J_i)a_{qf}(J_f)\tilde{b}_{1l}^{\nu}(J_i)\tilde{b}^{\nu}_{1q}(J_f)~_{\nu}'\langle J_f,1||T^{(E2)}||J_i,1\rangle'_{\nu}\Big|_{\nu=N+2} \nonumber\\
&\quad +\sum_{m,r} a_{mi}(J_i)a_{rf}(J_f)\tilde{b}_{1m}^{\nu}(J_i)\tilde{b}^{\nu}_{1r}(J_f)~_{\nu}'\langle J_f,1||T^{(E2)}||J_i,1\rangle'_{\nu}\Big|_{\nu=N+4}\nonumber\\
&= R(N+2)\langle J_f,f||T^{(E2)}||J_i,i\rangle + R(N+4)  \langle J_f,f||T^{(E2)}||J_i,i\rangle~,
\label{eq:contribution}
\end{align}
\end{widetext}
when $J_i(i)\rightarrow J_f(f)$ is an in-band transition in $^{186-192}$Pb. Thus, $R(N+2)$ gives the contribution in terms of percentage of the reduced matrix element of the $J_i(1)\rightarrow J_f(1)$ $E2$ transition in the lowest unperturbed 2p-2h band to the reduced matrix element of the $J_i(i)\rightarrow J_f(f)$ $E2$ transition in the fully correlated system. Similarly, $R(N+4)$ gives the contribution in terms of percentage of the reduced matrix element of the $J_i(1)\rightarrow J_f(1)$ $E2$ transition in the lowest unperturbed 4p-4h band. These ratios provide precise information on exactly which unperturbed states are mixed. We refer the reader to Ref. \cite{hellemans05} for a more detailed discussion.
\footnotetext[1]{in $J_i(i)$ and $J_f(f)$, the $i$ and $f$ between brackets are the ranknumbers of respectively the initial state $J_i$ and the final state $J_f$}

\begin{figure}[!t]
\begin{center}
\includegraphics[width=\columnwidth]{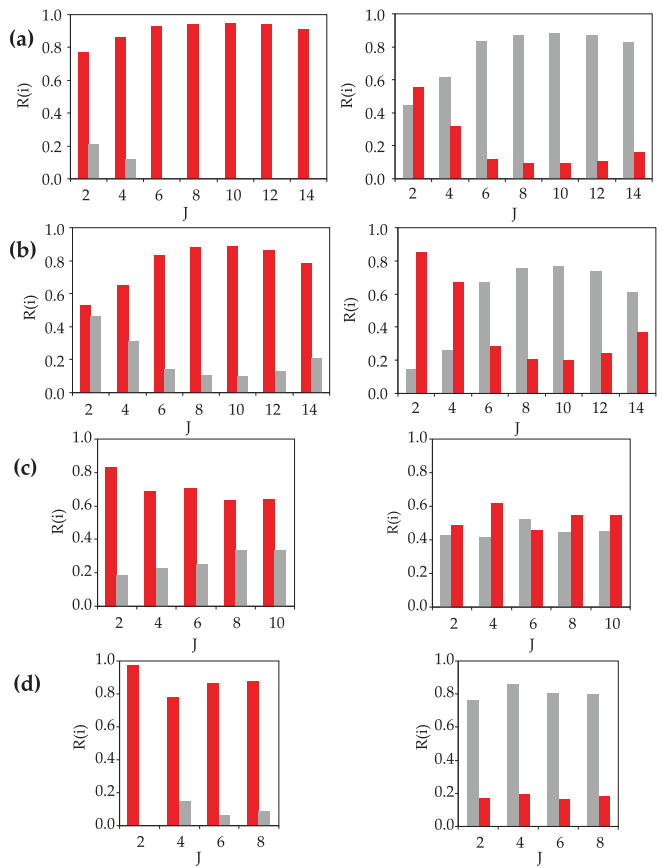}
\caption{(Color online) Ratios for the in-band transitions in (a)$^{186}$Pb (b) $^{188}$Pb, (c) $^{190}$Pb, and (d) $^{192}$Pb [see Eq. \eqref{eq:contribution}]. In the left (right) panels the ratios for the mainly 4p-4h (2p-2h) bands are shown. These correspond to Band I (Band II) in Ref. \cite{pakarinen07} for (a), to Band I (Band II) in Fig. \ref{fig:levelscheme188} for (b), and to Band II (Band I) in Figs. \ref{fig:levelscheme190/2} for (c) and (d). $R(N+2)$ is shown in gray whereas $R(N+4)$ is represented in red.}\label{fig:contribution}
\end{center}
\end{figure}

For $^{194-196}$Pb, transitions within the mixed 2p-2h band [Band I in Fig. \ref{fig:levelscheme194/6}(a) and \ref{fig:levelscheme194/6}(b)] originate for more than 90\% from the corresponding transition in the lowest unperturbed 2p-2h band. The transitions in the mixed 4p-4h states in these two isotopes contain important contributions from the second excited 2p-2h band.\\
The ratios $R(i)$ ($i=N+2,N+4$) for the mixed 2p-2h (right panels) and the mixed 4p-4h bands (left panels) in $^{186-192}$Pb are displayed in Fig. \ref{fig:contribution}. It follows that, in general, for the in-band $E2$ transitions in $^{186}$Pb and $^{192}$Pb the main contribution is coming from the $E2$ transition in the corresponding unperturbed band. In $^{190}$Pb, all states are very mixed, whereas in $^{186-188}$Pb, mainly the low-spin states are considerably more mixed. In particular, for the mixed 2p-2h bands in $^{186-188}$Pb [right panels of Figs. \ref{fig:contribution}(a) and \ref{fig:contribution}(b)] the contribution originating from the transition in  the unperturbed 4p-4h band exceeds the contribution of the transition in the unperturbed 2p-2h band for the low-spin states. This is due to large mixing and the larger quadrupole deformation of the 4p-4h band. This strong mixing is also reflected in the behavior of the quadrupole deformations (see Fig. \ref{fig:multiplot_iqm}).\\

\begin{figure*}[!tbp]
\begin{center}
 \includegraphics[width=0.9\textwidth]{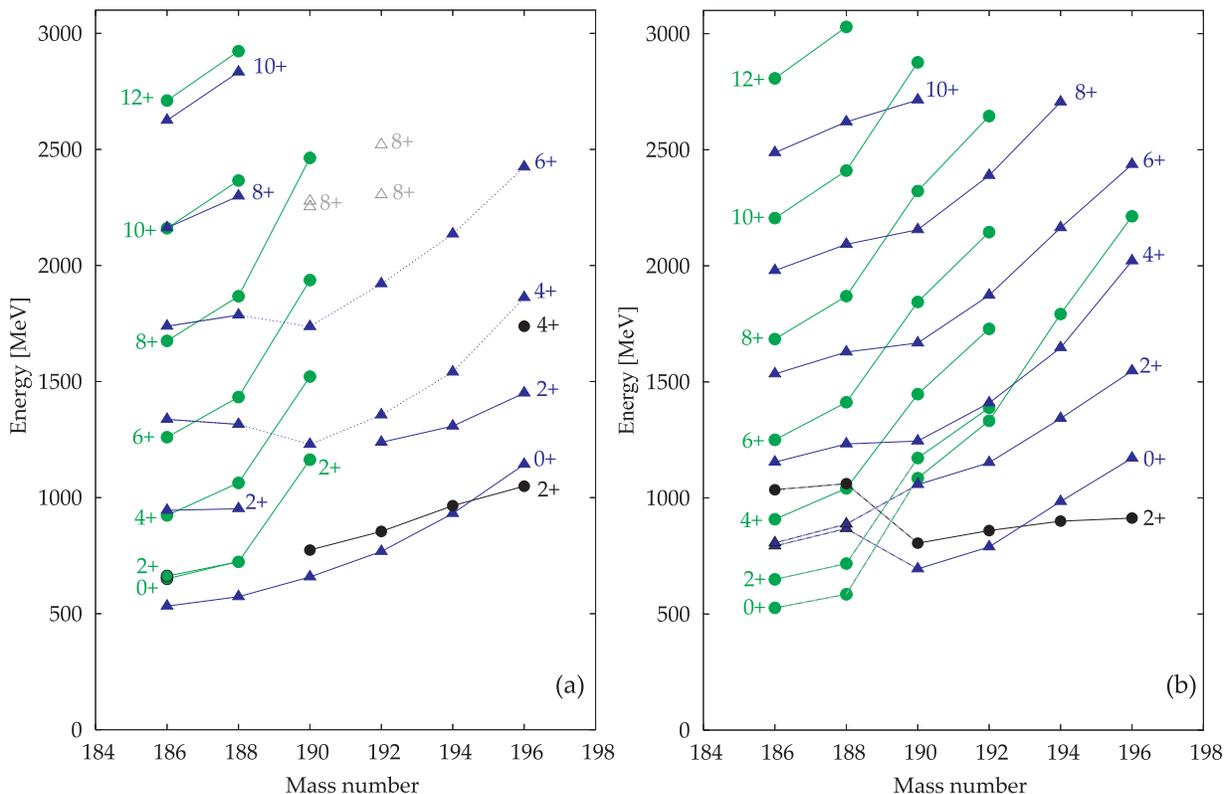}
\caption{(Color online) Experimental (a) and calculated (b) level systematics for $^{186-196}$Pb. In (b), states in black are mainly 0p-0h states, the states in blue (green) belong to the mainly 2p-2h (4p-4h) bands that were defined following the $E2$ decay starting from the high-spin states. For the dashed lines, see discussion in the text. In (a), the same color code is used but the states were organized on the basis of level systematics. The dotted line displays the suggested interpretation of the $4^+_1$ and $6^+_1$ states as possible 2p-2h states (see discussion). The $8^+$ states shown in gray are associated with spherical or oblate states \cite{pakarinen05b,julin01,dracoulis99,dracoulis98}. All states are shown relatively to the ground state. Experimental data were taken from Refs. \cite{pakarinen07,dracoulis04,dracoulis98,vanduppen87,plompen93,mcnabb97,dracoulis05,kaci02,penninga87,julin01,dracoulis99} and references therein.}\label{fig:systematics}
\end{center}
\end{figure*}

Summarizing, we present the experimental and the theoretically calculated level systematics for the $^{186-196}$Pb isotopes in Fig. \ref{fig:systematics}. In Fig. \ref{fig:systematics}(a) , the classification of the experimental states on the basis of level systematics \cite{pakarinen05b,julin01,dracoulis99,dracoulis98} is shown. The states in black are interpreted as 0p-0h states and the states in blue (green) were classified as 2p-2h (4p-4h) states. The $8^+$ states displayed in gray are associated with oblate or spherical states \cite{pakarinen05b,julin01,dracoulis99,dracoulis98}. Except for the $4^+$ and $6^+$ states connected with a dotted line, the systematics as proposed in \cite{pakarinen05b,julin01,dracoulis99,dracoulis98} were followed. For those $4^+$ and $6^+$ states in $^{190-194}$Pb that were previously assigned as spherical states, an alternative classification as states of 2p-2h character is proposed.
In Fig. \ref{fig:systematics}(b), the states in black are mainly 0p-0h states whereas the states in blue (green) belong to the mainly 2p-2h (4p-4h) bands that were defined following the $E2$ decay starting from high-spin states. The dashed lines indicate states that are very mixed. Although they belong to the mainly 2p-2h (4p-4h) bands defined by the $E2$ decay from higher-spin states, the weight of the corresponding wave function can be slightly higher in a different subspace (see Table \ref{tab:appendix} in Appendix \ref{appendix}). Hence, one should be prudent when classifying these highly mixed states as they can be organized following the $E2$ decay from high-spin states or according to the weight of the wave function in each subspace $[N]$, $[N+2]$, and $[N+4]$. These two criteria do not necessarily result in the same classification if the states are strongly mixed. A classification of the experimental states on the basis of $B(E2)$ values would resolve this ambiguity and provide a good basis for comparison.

\section{Conclusions}

In the present article, we have further explored the interacting boson model with configuration mixing for the neutron deficient Pb nuclei with 186$\le$A$\le$196. Proceeding from the more schematic calculations carried out by Fossion \textit{et al}. \cite{fossion03}, we have used adjusted parameters in the mixing Hamiltonian and have introduced a slight variation in the value of $\Delta^{N+2}$. With this parametrization, a successful description of the low-lying energy spectra, the in-band $B(E2)$ values and the interband $B(E2)$ ratios was achieved for the well-documented $^{186,188}$Pb isotopes. For an extensive discussion on the experimental data and the theoretical results on the midshell nucleus $^{186}$Pb, we refer the reader to Pakarinen \textit{et al.} \cite{pakarinen07}. The same parametrization was already used for the description of the isotopic shifts in $^{186-196}$Pb \cite{dewitte07}.

For the $^{186,188}$Pb isotopes, two low-lying collective bands have been measured up to high-spin states and were associated with 2p-2h and 4p-4h proton excitations across the closed $Z$=82 proton shell \cite{pakarinen07,dracoulis04}. If one proceeds  further away from midshell, less experimental evidence for well-developed collective bands is available. The available data display at most only one collective band and/or the onset of a collective band, which renders the interpretation of these states more difficult. On the basis of systematics \cite{pakarinen05b,julin01,dracoulis99}, the $4^+_1$ and the $6^+_1$ states in $^{190-194}$Pb were associated with spherical states. However, our present calculations display a strong $4^+_1\rightarrow 2^+_1$ transition and suggest that the former states might have a major, even dominant, 2p-2h character or are strongly mixed with the spherical $4^+_1$ and the $6^+_1$ states.

We have also extracted quadrupole deformations for the $^{186-196}$Pb nuclei. It is shown that within the assumption of a $K=0$ band structure for the lowest-lying bands in these nuclei, the unperturbed 2p-2h bands correspond to an oblate deformation whereas the unperturbed 4p-4h bands exhibit a prolate deformation. These results are consistent with the results of recent mean-field calculations \cite{bender04,rodriguez04} that display a prolate and and oblate minimum in the energy surface in addition to the spherical energy minimum. Our present calculations also show that mixing between the regular, 2p-2h and 4p-4h unperturbed configurations lowers the magnitude of the deformation. In the case of $^{190}$Pb, where the mixing is largest, this 
effect is most clearly visible.

Finally, we compared the experimental level systematics based on those presented in Refs. \cite{pakarinen05b,julin01,dracoulis99} with the members of the calculated low-lying bands for $^{186-196}$Pb. When confronting the theoretical to the experimental systematics, one should keep in mind that the theoretical states are classified according to the bands they belong to. These were defined following the $E2$ decay starting from the high-spin states. From an experimental point of view however, the states in $^{190-196}$Pb are mostly organized according to their assumed $n$p-$n$h character, based on the argument of level systematics. This classification becomes ambiguous when the states are very mixed or when little experimental data are available. Detailed in-beam spectroscopic data completed with lifetime measurements are therefore essential for a meaningful comparison between the experimental data and the theoretical results.

As a final remark, we point out that the IBM truncation provides a good description of the collective bands arising from 2p-2h and 4p-4h proton excitations across the $Z$=82 closed shell but is limited in the description of the spherical states that result from neutron pair breaking mainly. The spherical $4^+$, $6^+$, ... cannot be described adequately within the IBM with $s$ and $d$ bosons. Therefore, a truncated shell-model calculation that takes into account the open 82$\le$N$\le$126 neutron shell as well as a limited number of $n$p-$n$h proton excitations across the $Z$=82 closed shell is needed as a next step in the description of the rich structure of the neutron-deficient Pb isotopes from the neutron midshell region up to the doubly closed-shell nucleus $^{208}$Pb.

\begin{acknowledgments}
The authors thank R. Julin, J. Pakarinen, and P. Van Duppen for interesting discussions. Financial support from the ``FWO Vlaanderen'' (V.H. and K.H), the University of Ghent (S.D.B. and K.H.) and the Interuniversity Attraction Pole (IUAP) under projects P5/07 and P6/23 is gratefully acknowledged.
\end{acknowledgments}

\begin{appendix}
\section{Mixing amplitudes}\label{appendix}
In this appendix, we present the weight of the eigenvectors of the IBM Hamiltonian \eqref{eq:ibmhamiltonian} in the three subspaces $[N]$, $[N+2]$, and $[N+4]$ for the $^{186-196}$Pb isotopes.
\begin{table*}[!t]
\begin{center}
\caption{Weights of the IBM wave function in the different subspaces $[N]$, $[N+2]$, and $[N+4]$ for the $^{186-196}$Pb isotopes. Those weights smaller than 0.01 are denoted by 0.}\label{tab:appendix}
\begin{tabular}{c|ccc|ccc|ccc|ccc|ccc|ccc}
\hline\hline
$J(k)$ & \multicolumn{3}{c}{$^{186}$Pb} &\multicolumn{3}{c}{$^{188}$Pb}& \multicolumn{3}{c}{$^{190}$Pb} & \multicolumn{3}{c}{$^{192}$Pb}& \multicolumn{3}{c}{$^{194}$Pb}&\multicolumn{3}{c}{$^{196}$Pb}\\
\hline\hline
~ &$N$ & $N+2$ & $N+4$ & $N$ & $N+2$ & $N+4$ & $N$ & $N+2$ & $N+4$ & $N$ & $N+2$ & $N+4$ & $N$ & $N+2$ & $N+4$ & $N$ & $N+2$ & $N+4$\\
\hline
0(1)&	0.93&	0.06&	0&	0.93&	0.07&	0&	0.95&	0.05&	0&	0.96&	0.04&	0&	0.98&	0.02&	0&	0.99&	0.01&	0\\
0(2)&	0.03&	0.34&	0.63&	0.05&	0.55&	0.40&	0.05&	0.77&	0.18&	0.04&	0.86&	0.10&	0.03&	0.93&	0.05&	0.02&	0.96&	0.03\\
0(3)&	0.03&	0.58&	0.38&	0.02&	0.37&	0.61&	0.01&	0.18&	0.81&	0&	0.13&	0.87&	0.12&	0.07&	0.80&	0&	0.08&	0.91\\
2(1)&	0.07&	0.29&	0.64&	0.20&	0.50&	0.30&	0.47&	0.47&	0.07&	0.70&	0.29&	0.02&	0.91&	0.09&	0&	0.96&	0.04&	0\\
2(2)&	0.31&	0.37&	0.33&	0.26&	0.13&	0.61&	0.43&	0.20&	0.37&	0.29&	0.60&	0.11&	0.09&	0.86&	0.04&	0.04&	0.94&	0.02\\
2(3)&	0.54&	0.42&	0.03&	0.48&	0.47&	0.05&	0.09&	0.49&	0.42&	0.04&	0.50&	0.46&	0.25&	0.67&	0.08&	0.01&	0.24&	0.74\\
2(4)&	~&	~&	~&	~&	~&	~&	~&	~&	~&	~&	~&	~&	0.09&	0.25&	0.67&	~&	~&	~\\
4(1)&	0&	0.13&	0.87&	0.01&	0.28&	0.71&	0.03&	0.67&	0.29&	0.07&	0.83&	0.11&	0.22&	0.75&	0.03&	0.27&	0.71&	0.02\\
4(2)&	0.03&	0.84&	0.13&	0.03&	0.69&	0.28&	0.02&	0.32&	0.66&	0.02&	0.21&	0.77&	0.02&	0.62&	0.36&	0.36&	0.14&	0.51\\
4(3)&	 ~&	~&	~&	~&	~&	~&	~&	~&	~&	~&	~&	~&	0.02&	0.38&	0.60&	~&	~&	~\\
6(1)&	0&	0.09&	0.91&	0&	0.19&	0.81&	0.01&	0.57&	0.42&	0.01&	0.85&	0.13&	0.03&	0.94&	0.04&	0.06&	0.93&	0.02\\
6(2)&	0.01&	0.89&	0.10&	0.01&	0.81&	0.18&	0.01&	0.44&	0.55&	0&	0.19&	0.81&	0&	0.34&	0.66&	0&	0.12&	0.87\\
8(1)&	0&	0.08&	0.92&	0&	0.16&	0.84&	0&	0.53&	0.46&	0.01&	0.86&	0.13&	0.01&	0.96&	0.03&	~&	~&	~\\
8(2)&	0.01&	0.91&	0.08&	0.01&	0.84&	0.16&	0&	0.48&	0.52&	0&	0.17&	0.83&	0&	0.19&	0.81&	~&	~&	~\\
10(1)&	0&	0.08&	0.92&	0&	0.17&	0.83&	0&	0.58&	0.42&	~&	~&	~&	~&	~&	~&	~&	~&	~\\
10(2)&	0&	0.91&	0.08&	0&	0.83&	0.16&	0&	0.43&	0.57&	~&	~&	~&	~&	~&	~&	~&	~&	~\\
12(1)&	0&	0.10&	0.90&	0&	0.22&	0.78&	~&	~&	~&	~&	~&	~&	~&	~&	~&	~&	~&	~\\
12(2)&	0&	0.90&	0.10&	0&	0.78&	0.22&	~&	~&	~&	~&	~&	~&	~&	~&	~&	~&	~&	~\\
14(1)&	0&	0.15&	0.84&	0&	0.38&	0.62&	~&	~&	~&	~&	~&	~&	~&	~&	~&	~&	~&	~\\
14(2)&	0&	0.85&	0.15&	0&	0.63&	0.37&	~&	~&	~&	~&	~&	~&	~&	~&	~&	~&	~&	~\\
\hline\hline

\end{tabular}

\end{center}
\end{table*}

\end{appendix}

\bibliographystyle{apsrev}

\begin{thebibliography}{43}
\expandafter\ifx\csname natexlab\endcsname\relax\def\natexlab#1{#1}\fi
\expandafter\ifx\csname bibnamefont\endcsname\relax
  \def\bibnamefont#1{#1}\fi
\expandafter\ifx\csname bibfnamefont\endcsname\relax
  \def\bibfnamefont#1{#1}\fi
\expandafter\ifx\csname citenamefont\endcsname\relax
  \def\citenamefont#1{#1}\fi
\expandafter\ifx\csname url\endcsname\relax
  \def\url#1{\texttt{#1}}\fi
\expandafter\ifx\csname urlprefix\endcsname\relax\def\urlprefix{URL }\fi
\providecommand{\bibinfo}[2]{#2}
\providecommand{\eprint}[2][]{\url{#2}}

\bibitem[{\citenamefont{Wood et~al.}(1992)\citenamefont{Wood, Heyde,
  Nazarewicz, Huyse, and Van~Duppen}}]{wood92}
\bibinfo{author}{\bibfnamefont{J.~L.} \bibnamefont{Wood}},
  \bibinfo{author}{\bibfnamefont{K.}~\bibnamefont{Heyde}},
  \bibinfo{author}{\bibfnamefont{W.}~\bibnamefont{Nazarewicz}},
  \bibinfo{author}{\bibfnamefont{M.}~\bibnamefont{Huyse}}, \bibnamefont{and}
  \bibinfo{author}{\bibfnamefont{P.}~\bibnamefont{Van~Duppen}},
  \bibinfo{journal}{Phys. Rep.} \textbf{\bibinfo{volume}{215}},
  \bibinfo{pages}{101} (\bibinfo{year}{1992}).

\bibitem[{\citenamefont{Andreyev et~al.}(2000)\citenamefont{Andreyev, Huyse,
  Van~Duppen, Weismann, Ackermann, Gerl, Heissberger, Hoffmann, Kleinbohl,
  Munzenberger et~al.}}]{andreyev00}
\bibinfo{author}{\bibfnamefont{A.~N.}~\bibnamefont{Andreyev}},
  \bibinfo{author}{\bibfnamefont{M.}~\bibnamefont{Huyse}},
  \bibinfo{author}{\bibfnamefont{P.} \bibnamefont{Van~Duppen}},
  \bibinfo{author}{\bibfnamefont{L.}~\bibnamefont{Weissmann}},
  \bibinfo{author}{\bibfnamefont{D.}~\bibnamefont{Ackermann}},
  \bibinfo{author}{\bibfnamefont{J.}~\bibnamefont{Gerl}},
  \bibinfo{author}{\bibfnamefont{F.~P.}~\bibnamefont{Hessberger}},
  \bibinfo{author}{\bibfnamefont{S.}~\bibnamefont{Hoffmann}},
  \bibinfo{author}{\bibfnamefont{A.}~\bibnamefont{Kleinb\"ohl}},
  \bibinfo{author}{\bibfnamefont{G.}~\bibnamefont{M\"unzenberg}}
  \bibnamefont{\textit{et~al.}}, \bibinfo{journal}{Nature}
  \textbf{\bibinfo{volume}{405}}, \bibinfo{pages}{430} (\bibinfo{year}{2000}).

\bibitem[{\citenamefont{Julin et~al.}(2001)\citenamefont{Julin, Helariutta, and
  Muikku}}]{julin01}
\bibinfo{author}{\bibfnamefont{R.}~\bibnamefont{Julin}},
  \bibinfo{author}{\bibfnamefont{K.}~\bibnamefont{Helariutta}},
  \bibnamefont{and} \bibinfo{author}{\bibfnamefont{M.}~\bibnamefont{Muikku}},
  \bibinfo{journal}{J. Phys. G} \textbf{\bibinfo{volume}{27}},
  \bibinfo{pages}{R109} (\bibinfo{year}{2001}).

\bibitem[{\citenamefont{Van~Duppen et~al.}(1984)\citenamefont{Van~Duppen,
  Coenen, Deneffe, Huyse, Heyde, and Van~Isacker}}]{vanduppen84}
\bibinfo{author}{\bibfnamefont{P.}~\bibnamefont{Van~Duppen}},
  \bibinfo{author}{\bibfnamefont{E.}~\bibnamefont{Coenen}},
  \bibinfo{author}{\bibfnamefont{K.}~\bibnamefont{Deneffe}},
  \bibinfo{author}{\bibfnamefont{M.}~\bibnamefont{Huyse}},
  \bibinfo{author}{\bibfnamefont{K.}~\bibnamefont{Heyde}}, \bibnamefont{and}
  \bibinfo{author}{\bibfnamefont{P.}~\bibnamefont{Van~Isacker}},
  \bibinfo{journal}{Phys. Rev. Lett.} \textbf{\bibinfo{volume}{52}},
  \bibinfo{pages}{1974} (\bibinfo{year}{1984}).

\bibitem[{\citenamefont{De~Coster et~al.}(2000)\citenamefont{De~Coster,
  Decroix, and Heyde}}]{coster00}
\bibinfo{author}{\bibfnamefont{C.}~\bibnamefont{De~Coster}},
  \bibinfo{author}{\bibfnamefont{B.}~\bibnamefont{Decroix}}, \bibnamefont{and}
  \bibinfo{author}{\bibfnamefont{K.}~\bibnamefont{Heyde}},
  \bibinfo{journal}{Phys. Rev. C} \textbf{\bibinfo{volume}{61}},
  \bibinfo{pages}{067306} (\bibinfo{year}{2000}).

\bibitem[{\citenamefont{Heyde et~al.}(1987)\citenamefont{Heyde, Jolie, Moreau,
  Ryckebusch, Waroquier, Van~Duppen, Huyse, and Wood}}]{heyde87}
\bibinfo{author}{\bibfnamefont{K.}~\bibnamefont{Heyde}},
  \bibinfo{author}{\bibfnamefont{J.}~\bibnamefont{Jolie}},
  \bibinfo{author}{\bibfnamefont{J.}~\bibnamefont{Moreau}},
  \bibinfo{author}{\bibfnamefont{J.}~\bibnamefont{Ryckebusch}},
  \bibinfo{author}{\bibfnamefont{M.}~\bibnamefont{Waroquier}},
  \bibinfo{author}{\bibfnamefont{P.}~\bibnamefont{Van~Duppen}},
  \bibinfo{author}{\bibfnamefont{M.}~\bibnamefont{Huyse}}, \bibnamefont{and}
  \bibinfo{author}{\bibfnamefont{J.~L.} \bibnamefont{Wood}},
  \bibinfo{journal}{Nucl. Phys. A} \textbf{\bibinfo{volume}{466}},
  \bibinfo{pages}{189} (\bibinfo{year}{1987}).

\bibitem[{\citenamefont{Witte et~al.}(2007)\citenamefont{Witte, Andreyev,
  Barre, Bender, Cocolios, Dean, Fedorov, Fedoseyev, Fraile, Franchoo
  et~al.}}]{dewitte07}
\bibinfo{author}{\bibfnamefont{H.} \bibnamefont{De~Witte}},
  \bibinfo{author}{\bibfnamefont{A.~N.} \bibnamefont{Andreyev}},
  \bibinfo{author}{\bibfnamefont{N.}~\bibnamefont{Barre}},
  \bibinfo{author}{\bibfnamefont{M.}~\bibnamefont{Bender}},
  \bibinfo{author}{\bibfnamefont{T.~E.} \bibnamefont{Cocolios}},
  \bibinfo{author}{\bibfnamefont{S.}~\bibnamefont{Dean}},
  \bibinfo{author}{\bibfnamefont{D.}~\bibnamefont{Fedorov}},
  \bibinfo{author}{\bibfnamefont{V.~N.} \bibnamefont{Fedoseyev}},
  \bibinfo{author}{\bibfnamefont{L.~M.} \bibnamefont{Fraile}},
  \bibinfo{author}{\bibfnamefont{S.}~\bibnamefont{Franchoo}}
  \bibnamefont{\textit{et~al.}}, \bibinfo{journal}{Phys. Rev. Lett.}
  \textbf{\bibinfo{volume}{98}}, \bibinfo{eid}{112502} (\bibinfo{year}{2007}).

\bibitem[{\citenamefont{Pakarinen et~al.}(2007)\citenamefont{Pakarinen,
  Hellemans, Julin, Juutinen, Heyde, Heenen, Bender, Darby, Eeckhaudt, Enqvist
  et~al.}}]{pakarinen07}
\bibinfo{author}{\bibfnamefont{J.}~\bibnamefont{Pakarinen}},
  \bibinfo{author}{\bibfnamefont{V.}~\bibnamefont{Hellemans}},
  \bibinfo{author}{\bibfnamefont{R.}~\bibnamefont{Julin}},
  \bibinfo{author}{\bibfnamefont{S.}~\bibnamefont{Juutinen}},
  \bibinfo{author}{\bibfnamefont{K.}~\bibnamefont{Heyde}},
  \bibinfo{author}{\bibfnamefont{P.-H.} \bibnamefont{Heenen}},
  \bibinfo{author}{\bibfnamefont{M.}~\bibnamefont{Bender}},
  \bibinfo{author}{\bibfnamefont{I.~G.} \bibnamefont{Darby}},
  \bibinfo{author}{\bibfnamefont{S.}~\bibnamefont{Eeckhaudt}},
  \bibinfo{author}{\bibfnamefont{T.}~\bibnamefont{Enqvist}}
  \bibnamefont{\textit{et~al.}}, \bibinfo{journal}{Phys. Rev. C}
  \textbf{\bibinfo{volume}{75}}, \bibinfo{eid}{014302} (\bibinfo{year}{2007}).

\bibitem[{\citenamefont{Grahn et~al.}(2006)\citenamefont{Grahn, Dewald, Moller,
  Julin, Beausang, Christen, Darby, Eeckhaudt, Greenlees, Gorgen
  et~al.}}]{grahn06}
\bibinfo{author}{\bibfnamefont{T.}~\bibnamefont{Grahn}},
  \bibinfo{author}{\bibfnamefont{A.}~\bibnamefont{Dewald}},
  \bibinfo{author}{\bibfnamefont{O.}~\bibnamefont{Moller}},
  \bibinfo{author}{\bibfnamefont{R.}~\bibnamefont{Julin}},
  \bibinfo{author}{\bibfnamefont{C.~W.} \bibnamefont{Beausang}},
  \bibinfo{author}{\bibfnamefont{S.}~\bibnamefont{Christen}},
  \bibinfo{author}{\bibfnamefont{I.~G.} \bibnamefont{Darby}},
  \bibinfo{author}{\bibfnamefont{S.}~\bibnamefont{Eeckhaudt}},
  \bibinfo{author}{\bibfnamefont{P.~T.} \bibnamefont{Greenlees}},
  \bibinfo{author}{\bibfnamefont{A.}~\bibnamefont{Gorgen}}
  \bibnamefont{\textit{et~al.}}, \bibinfo{journal}{Phys. Rev. Lett.}
  \textbf{\bibinfo{volume}{97}}, \bibinfo{eid}{062501} (\bibinfo{year}{2006}).

\bibitem[{\citenamefont{Dracoulis et~al.}(2005)\citenamefont{Dracoulis, Lane,
  Peaty, Byrne, Baxter, Davidson, Wilson, Kibedi, and Xu}}]{dracoulis05}
\bibinfo{author}{\bibfnamefont{G.~D.} \bibnamefont{Dracoulis}},
  \bibinfo{author}{\bibfnamefont{G.~J.} \bibnamefont{Lane}},
  \bibinfo{author}{\bibfnamefont{T.~M.} \bibnamefont{Peaty}},
  \bibinfo{author}{\bibfnamefont{A.~P.} \bibnamefont{Byrne}},
  \bibinfo{author}{\bibfnamefont{A.~M.} \bibnamefont{Baxter}},
  \bibinfo{author}{\bibfnamefont{P.~M.} \bibnamefont{Davidson}},
  \bibinfo{author}{\bibfnamefont{A.~N.} \bibnamefont{Wilson}},
  \bibinfo{author}{\bibfnamefont{T.}~\bibnamefont{Kibedi}}, \bibnamefont{and}
  \bibinfo{author}{\bibfnamefont{F.~R.} \bibnamefont{Xu}},
  \bibinfo{journal}{Phys. Rev. C} \textbf{\bibinfo{volume}{72}},
  \bibinfo{eid}{064319} (\bibinfo{year}{2005}).

\bibitem[{\citenamefont{Dracoulis et~al.}(2004)\citenamefont{Dracoulis, Lane,
  Byrne, Kibedi, Baxter, Macchiavelli, Fallon, and Clark}}]{dracoulis04}
\bibinfo{author}{\bibfnamefont{G.~D.} \bibnamefont{Dracoulis}},
  \bibinfo{author}{\bibfnamefont{G.~J.} \bibnamefont{Lane}},
  \bibinfo{author}{\bibfnamefont{A.~P.} \bibnamefont{Byrne}},
  \bibinfo{author}{\bibfnamefont{T.}~\bibnamefont{Kibedi}},
  \bibinfo{author}{\bibfnamefont{A.~M.} \bibnamefont{Baxter}},
  \bibinfo{author}{\bibfnamefont{A.~O.} \bibnamefont{Macchiavelli}},
  \bibinfo{author}{\bibfnamefont{P.}~\bibnamefont{Fallon}}, \bibnamefont{and}
  \bibinfo{author}{\bibfnamefont{R.~M.} \bibnamefont{Clark}},
  \bibinfo{journal}{Phys. Rev. C} \textbf{\bibinfo{volume}{69}},
  \bibinfo{eid}{054318} (\bibinfo{year}{2004}).

\bibitem[{\citenamefont{Ionescu-Bujor et~al.}(2007)\citenamefont{Ionescu-Bujor,
  Iordachescu, Marginean, Ur, Bucurescu, Suliman, Balabanski, Brandolini,
  Chmel, Detistov et~al.}}]{bujor07}
\bibinfo{author}{\bibfnamefont{M.}~\bibnamefont{Ionescu-Bujor}},
  \bibinfo{author}{\bibfnamefont{A.}~\bibnamefont{Iordachescu}},
  \bibinfo{author}{\bibfnamefont{N.}~\bibnamefont{Marginean}},
  \bibinfo{author}{\bibfnamefont{C.~A.} \bibnamefont{Ur}},
  \bibinfo{author}{\bibfnamefont{D.}~\bibnamefont{Bucurescu}},
  \bibinfo{author}{\bibfnamefont{G.}~\bibnamefont{Suliman}},
  \bibinfo{author}{\bibfnamefont{D.~L.} \bibnamefont{Balabanski}},
  \bibinfo{author}{\bibfnamefont{F.}~\bibnamefont{Brandolini}},
  \bibinfo{author}{\bibfnamefont{S.}~\bibnamefont{Chmel}},
  \bibinfo{author}{\bibfnamefont{P.}~\bibnamefont{Detistov}}
  \bibnamefont{\textit{et~al.}}, \bibinfo{journal}{Phys. Lett. B}
  \textbf{\bibinfo{volume}{650}}, \bibinfo{pages}{141} (\bibinfo{year}{2007}).

\bibitem[{\citenamefont{Dewald et~al.}(2003)\citenamefont{Dewald, Peusquens,
  Saha, von Brentano, Fitzler, Klug, Wiedenh\"over, Carpenter, Heinz, Janssens
  et~al.}}]{dewald03}
\bibinfo{author}{\bibfnamefont{A.}~\bibnamefont{Dewald}},
  \bibinfo{author}{\bibfnamefont{R.}~\bibnamefont{Peusquens}},
  \bibinfo{author}{\bibfnamefont{B.}~\bibnamefont{Saha}},
  \bibinfo{author}{\bibfnamefont{P.}~\bibnamefont{von Brentano}},
  \bibinfo{author}{\bibfnamefont{A.}~\bibnamefont{Fitzler}},
  \bibinfo{author}{\bibfnamefont{T.}~\bibnamefont{Klug}},
  \bibinfo{author}{\bibfnamefont{I.}~\bibnamefont{Wiedenh\"over}},
  \bibinfo{author}{\bibfnamefont{M.}~\bibnamefont{Carpenter}},
  \bibinfo{author}{\bibfnamefont{A.}~\bibnamefont{Heinz}},
  \bibinfo{author}{\bibfnamefont{R.~V.~F.} \bibnamefont{Janssens}}
  \bibnamefont{\textit{et~al.}}, \bibinfo{journal}{Phys. Rev. C}
  \textbf{\bibinfo{volume}{68}}, \bibinfo{pages}{034314}
  (\bibinfo{year}{2003}).

\bibitem[{\citenamefont{May et~al.}(1977)\citenamefont{May, Pashkevich, and
  Frauendorf}}]{may77}
\bibinfo{author}{\bibfnamefont{F.~R.} \bibnamefont{May}},
  \bibinfo{author}{\bibfnamefont{V.~V.} \bibnamefont{Pashkevich}},
  \bibnamefont{and}
  \bibinfo{author}{\bibfnamefont{S.}~\bibnamefont{Frauendorf}},
  \bibinfo{journal}{Phys. Lett. B} \textbf{\bibinfo{volume}{68}},
  \bibinfo{pages}{113} (\bibinfo{year}{1977}).

\bibitem[{\citenamefont{Nazarewicz}(1993)}]{nazarewicz93}
\bibinfo{author}{\bibfnamefont{W.}~\bibnamefont{Nazarewicz}},
  \bibinfo{journal}{Phys. Lett. B} \textbf{\bibinfo{volume}{305}},
  \bibinfo{pages}{195} (\bibinfo{year}{1993}).

\bibitem[{\citenamefont{Tajima et~al.}(1993)\citenamefont{Tajima, Flocard,
  Bonche, Dobaczewski, and Heenen}}]{tajima93}
\bibinfo{author}{\bibfnamefont{N.}~\bibnamefont{Tajima}},
  \bibinfo{author}{\bibfnamefont{H.}~\bibnamefont{Flocard}},
  \bibinfo{author}{\bibfnamefont{P.}~\bibnamefont{Bonche}},
  \bibinfo{author}{\bibfnamefont{J.}~\bibnamefont{Dobaczewski}},
  \bibnamefont{and} \bibinfo{author}{\bibfnamefont{P.~H.}
  \bibnamefont{Heenen}}, \bibinfo{journal}{Nucl. Phys. A}
  \textbf{\bibinfo{volume}{551}}, \bibinfo{pages}{409} (\bibinfo{year}{1993}).

\bibitem[{\citenamefont{Bengsston and Nazarewicz}(1989)}]{bengsston89}
\bibinfo{author}{\bibfnamefont{R.}~\bibnamefont{Bengsston}} \bibnamefont{and}
  \bibinfo{author}{\bibfnamefont{W.}~\bibnamefont{Nazarewicz}},
  \bibinfo{journal}{Z. Phys. A} \textbf{\bibinfo{volume}{334}},
  \bibinfo{pages}{269} (\bibinfo{year}{1989}).

\bibitem[{\citenamefont{Bender et~al.}(2004)\citenamefont{Bender, Bonche,
  Duguet, and Heenen}}]{bender04}
\bibinfo{author}{\bibfnamefont{M.}~\bibnamefont{Bender}},
  \bibinfo{author}{\bibfnamefont{P.}~\bibnamefont{Bonche}},
  \bibinfo{author}{\bibfnamefont{T.}~\bibnamefont{Duguet}}, \bibnamefont{and}
  \bibinfo{author}{\bibfnamefont{P.-H.} \bibnamefont{Heenen}},
  \bibinfo{journal}{Phys. Rev. C} \textbf{\bibinfo{volume}{69}},
  \bibinfo{pages}{064303} (\bibinfo{year}{2004}).

\bibitem[{\citenamefont{Rodriguez{-}Guzman
  et~al.}(2004)\citenamefont{Rodriguez{-}Guzman, Egido, and
  Robledo}}]{rodriguez04}
\bibinfo{author}{\bibfnamefont{R.~R.} \bibnamefont{Rodriguez{-}Guzman}},
  \bibinfo{author}{\bibfnamefont{J.~L.} \bibnamefont{Egido}}, \bibnamefont{and}
  \bibinfo{author}{\bibfnamefont{L.~M.} \bibnamefont{Robledo}},
  \bibinfo{journal}{Phys. Rev. C} \textbf{\bibinfo{volume}{69}},
  \bibinfo{eid}{054319} (\bibinfo{year}{2004}).

\bibitem[{\citenamefont{Iachello and Arima}(1987)}]{iachello87}
\bibinfo{author}{\bibfnamefont{F.}~\bibnamefont{Iachello}} \bibnamefont{and}
  \bibinfo{author}{\bibfnamefont{A.}~\bibnamefont{Arima}},
  \emph{\bibinfo{title}{The Interacting Boson Model}}
  (\bibinfo{publisher}{Cambridge University Press},
  \bibinfo{address}{Cambridghe}, \bibinfo{year}{1987}).

\bibitem[{\citenamefont{Frank and Van~Isacker}(1994)}]{frank94}
\bibinfo{author}{\bibfnamefont{A.}~\bibnamefont{Frank}} \bibnamefont{and}
  \bibinfo{author}{\bibfnamefont{P.}~\bibnamefont{Van~Isacker}},
  \emph{\bibinfo{title}{Algebraic Methods in Molecular and Nuclear Structure
  Physics}} (\bibinfo{publisher}{John Wiley and Sons, Inc.},
  \bibinfo{address}{New York}, \bibinfo{year}{1994}).

\bibitem[{\citenamefont{Duval and Barrett}(1981)}]{duval81}
\bibinfo{author}{\bibfnamefont{P.~D.} \bibnamefont{Duval}} \bibnamefont{and}
  \bibinfo{author}{\bibfnamefont{B.~R.} \bibnamefont{Barrett}},
  \bibinfo{journal}{Phys. Lett. B} \textbf{\bibinfo{volume}{100}},
  \bibinfo{pages}{223} (\bibinfo{year}{1981}).

\bibitem[{\citenamefont{Duval and Barrett}(1982)}]{duval82}
\bibinfo{author}{\bibfnamefont{P.~D.} \bibnamefont{Duval}} \bibnamefont{and}
  \bibinfo{author}{\bibfnamefont{B.~R.} \bibnamefont{Barrett}},
  \bibinfo{journal}{Nucl. Phys. A} \textbf{\bibinfo{volume}{376}},
  \bibinfo{pages}{213} (\bibinfo{year}{1982}).

\bibitem[{\citenamefont{Hellemans et~al.}(2005)\citenamefont{Hellemans,
  Fossion, De~Baerdemacker, and Heyde}}]{hellemans05}
\bibinfo{author}{\bibfnamefont{V.}~\bibnamefont{Hellemans}},
  \bibinfo{author}{\bibfnamefont{R.}~\bibnamefont{Fossion}},
  \bibinfo{author}{\bibfnamefont{S.} \bibnamefont{De~Baerdemacker}},
  \bibnamefont{and} \bibinfo{author}{\bibfnamefont{K.}~\bibnamefont{Heyde}},
  \bibinfo{journal}{Phys. Rev. C} \textbf{\bibinfo{volume}{71}},
  \bibinfo{eid}{034308} (\bibinfo{year}{2005}).

\bibitem[{\citenamefont{Coster et~al.}(1996)\citenamefont{Coster, Heyde,
  Decroix, Van~Isacker, Jolie, Lehmann, and Wood}}]{coster96}
\bibinfo{author}{\bibfnamefont{C.~De} \bibnamefont{Coster}},
  \bibinfo{author}{\bibfnamefont{K.}~\bibnamefont{Heyde}},
  \bibinfo{author}{\bibfnamefont{B.}~\bibnamefont{Decroix}},
  \bibinfo{author}{\bibfnamefont{P.}~\bibnamefont{Van~Isacker}},
  \bibinfo{author}{\bibfnamefont{J.}~\bibnamefont{Jolie}},
  \bibinfo{author}{\bibfnamefont{H.}~\bibnamefont{Lehmann}}, \bibnamefont{and}
  \bibinfo{author}{\bibfnamefont{J.}~\bibnamefont{Wood}},
  \bibinfo{journal}{Nucl. Phys. A} \textbf{\bibinfo{volume}{600}},
  \bibinfo{pages}{251} (\bibinfo{year}{1996}).

\bibitem[{\citenamefont{Heyde et~al.}(1992)\citenamefont{Heyde, De~Coster,
  Jolie, and Wood}}]{heyde92}
\bibinfo{author}{\bibfnamefont{K.}~\bibnamefont{Heyde}},
  \bibinfo{author}{\bibfnamefont{C.}~\bibnamefont{De~Coster}},
  \bibinfo{author}{\bibfnamefont{J.}~\bibnamefont{Jolie}}, \bibnamefont{and}
  \bibinfo{author}{\bibfnamefont{J.~L.} \bibnamefont{Wood}},
  \bibinfo{journal}{Phys. Rev. C} \textbf{\bibinfo{volume}{46}},
  \bibinfo{pages}{541} (\bibinfo{year}{1992}).

\bibitem[{\citenamefont{Warner and Casten}(1983)}]{warner83}
\bibinfo{author}{\bibfnamefont{D.~D.} \bibnamefont{Warner}} \bibnamefont{and}
  \bibinfo{author}{\bibfnamefont{R.~F.} \bibnamefont{Casten}},
  \bibinfo{journal}{Phys. Rev. C} \textbf{\bibinfo{volume}{28}},
  \bibinfo{pages}{1798} (\bibinfo{year}{1983}).

\bibitem[{\citenamefont{Fossion et~al.}(2003)\citenamefont{Fossion, Heyde,
  Thiamova, and Van~Isacker}}]{fossion03}
\bibinfo{author}{\bibfnamefont{R.}~\bibnamefont{Fossion}},
  \bibinfo{author}{\bibfnamefont{K.}~\bibnamefont{Heyde}},
  \bibinfo{author}{\bibfnamefont{G.}~\bibnamefont{Thiamova}}, \bibnamefont{and}
  \bibinfo{author}{\bibfnamefont{P.}~\bibnamefont{Van~Isacker}},
  \bibinfo{journal}{Phys. Rev. C} \textbf{\bibinfo{volume}{67}},
  \bibinfo{pages}{024306} (\bibinfo{year}{2003}).

\bibitem[{\citenamefont{Heyde et~al.}(1991)\citenamefont{Heyde, Schietse, and
  De~Coster}}]{heyde91}
\bibinfo{author}{\bibfnamefont{K.}~\bibnamefont{Heyde}},
  \bibinfo{author}{\bibfnamefont{J.}~\bibnamefont{Schietse}}, \bibnamefont{and}
  \bibinfo{author}{\bibfnamefont{C.}~\bibnamefont{De~Coster}},
  \bibinfo{journal}{Phys. Rev. C} \textbf{\bibinfo{volume}{44}},
  \bibinfo{pages}{2216} (\bibinfo{year}{1991}).

\bibitem[{\citenamefont{Dracoulis et~al.}(1998)\citenamefont{Dracoulis, Byrne,
  and Baxter}}]{dracoulis98}
\bibinfo{author}{\bibfnamefont{G.~D.} \bibnamefont{Dracoulis}},
  \bibinfo{author}{\bibfnamefont{A.~P.} \bibnamefont{Byrne}}, \bibnamefont{and}
  \bibinfo{author}{\bibfnamefont{A.~M.} \bibnamefont{Baxter}},
  \bibinfo{journal}{Phys. Lett. B} \textbf{\bibinfo{volume}{432}},
  \bibinfo{pages}{37} (\bibinfo{year}{1998}).

\bibitem[{\citenamefont{Van~Duppen et~al.}(1987)\citenamefont{Van~Duppen,
  Coenen, Deneffe, Huyse, and Wood}}]{vanduppen87}
\bibinfo{author}{\bibfnamefont{P.}~\bibnamefont{Van~Duppen}},
  \bibinfo{author}{\bibfnamefont{E.}~\bibnamefont{Coenen}},
  \bibinfo{author}{\bibfnamefont{K.}~\bibnamefont{Deneffe}},
  \bibinfo{author}{\bibfnamefont{M.}~\bibnamefont{Huyse}}, \bibnamefont{and}
  \bibinfo{author}{\bibfnamefont{J.~L.} \bibnamefont{Wood}},
  \bibinfo{journal}{Phys. Rev. C} \textbf{\bibinfo{volume}{35}},
  \bibinfo{pages}{1861} (\bibinfo{year}{1987}).

\bibitem[{\citenamefont{Plompen et~al.}(1993)\citenamefont{Plompen, Harakeh,
  Hesselink, van~'t Hof, Kalantar-Nayestanaki, van Schagen, Carpenter, Ahmad,
  Bearden, Janssens et~al.}}]{plompen93}
\bibinfo{author}{\bibfnamefont{A.~J.~M.} \bibnamefont{Plompen}},
  \bibinfo{author}{\bibfnamefont{M.~N.} \bibnamefont{Harakeh}},
  \bibinfo{author}{\bibfnamefont{W.~H.~A.} \bibnamefont{Hesselink}},
  \bibinfo{author}{\bibfnamefont{G.}~\bibnamefont{van~'t Hof}},
  \bibinfo{author}{\bibfnamefont{N.}~\bibnamefont{Kalantar-Nayestanaki}},
  \bibinfo{author}{\bibfnamefont{J.~P.~S.} \bibnamefont{van Schagen}},
  \bibinfo{author}{\bibfnamefont{M.~P.} \bibnamefont{Carpenter}},
  \bibinfo{author}{\bibfnamefont{I.}~\bibnamefont{Ahmad}},
  \bibinfo{author}{\bibfnamefont{I.~G.} \bibnamefont{Bearden}},
  \bibinfo{author}{\bibfnamefont{R.~V.~F.} \bibnamefont{Janssens}}
  \bibnamefont{\textit{et~al.}}, \bibinfo{journal}{Nucl. Phys. A}
  \textbf{\bibinfo{volume}{562}}, \bibinfo{pages}{61} (\bibinfo{year}{1993}).

\bibitem[{\citenamefont{McNabb et~al.}(1997)\citenamefont{McNabb, Cizewski,
  Ding, Fotiades, Archer, Becker, Bernstein, Hauschild, Younes, Clark
  et~al.}}]{mcnabb97}
\bibinfo{author}{\bibfnamefont{D.~P.} \bibnamefont{McNabb}},
  \bibinfo{author}{\bibfnamefont{J.~A.} \bibnamefont{Cizewski}},
  \bibinfo{author}{\bibfnamefont{K.~Y.} \bibnamefont{Ding}},
  \bibinfo{author}{\bibfnamefont{N.}~\bibnamefont{Fotiades}},
  \bibinfo{author}{\bibfnamefont{D.~E.} \bibnamefont{Archer}},
  \bibinfo{author}{\bibfnamefont{J.~A.} \bibnamefont{Becker}},
  \bibinfo{author}{\bibfnamefont{L.~A.} \bibnamefont{Bernstein}},
  \bibinfo{author}{\bibfnamefont{K.}~\bibnamefont{Hauschild}},
  \bibinfo{author}{\bibfnamefont{W.}~\bibnamefont{Younes}},
  \bibinfo{author}{\bibfnamefont{R.~M.} \bibnamefont{Clark}}
  \bibnamefont{\textit{et~al.}}, \bibinfo{journal}{Phys. Rev. C}
  \textbf{\bibinfo{volume}{56}}, \bibinfo{pages}{2474} (\bibinfo{year}{1997}).

\bibitem[{\citenamefont{Kaci et~al.}(2002)\citenamefont{Kaci, Porquet,
  Deloncle, Aiche, Azaiaz, Bastin, Beausang, Bourgeois, Clark, Duffait
  et~al.}}]{kaci02}
\bibinfo{author}{\bibfnamefont{M.}~\bibnamefont{Kaci}},
  \bibinfo{author}{\bibfnamefont{M.-G.} \bibnamefont{Porquet}},
  \bibinfo{author}{\bibfnamefont{I.}~\bibnamefont{Deloncle}},
  \bibinfo{author}{\bibfnamefont{M.}~\bibnamefont{Aiche}},
  \bibinfo{author}{\bibfnamefont{F.}~\bibnamefont{Azaiaz}},
  \bibinfo{author}{\bibfnamefont{G.}~\bibnamefont{Bastin}},
  \bibinfo{author}{\bibfnamefont{C.~W.} \bibnamefont{Beausang}},
  \bibinfo{author}{\bibfnamefont{C.}~\bibnamefont{Bourgeois}},
  \bibinfo{author}{\bibfnamefont{R.~M.} \bibnamefont{Clark}},
  \bibinfo{author}{\bibfnamefont{R.}~\bibnamefont{Duffait}}
  \bibnamefont{\textit{et~al.}}, \bibinfo{journal}{Nucl. Phys. A}
  \textbf{\bibinfo{volume}{697}}, \bibinfo{pages}{3} (\bibinfo{year}{2002}).

\bibitem[{\citenamefont{Penninga and Hesselink}(1987)}]{penninga87}
\bibinfo{author}{\bibfnamefont{J.}~\bibnamefont{Penninga}} \bibnamefont{and}
  \bibinfo{author}{\bibfnamefont{W.~H.~A.} \bibnamefont{Hesselink}},
  \bibinfo{journal}{Nucl. Phys. A} \textbf{\bibinfo{volume}{471}},
  \bibinfo{pages}{535} (\bibinfo{year}{1987}).

\bibitem[{\citenamefont{Dendooven et~al.}(1989)\citenamefont{Dendooven,
  Decrock, Huysse, Reusen, Van~Duppen, and Wauters}}]{dendooven89}
\bibinfo{author}{\bibfnamefont{P.}~\bibnamefont{Dendooven}},
  \bibinfo{author}{\bibfnamefont{P.}~\bibnamefont{Decrock}},
  \bibinfo{author}{\bibfnamefont{M.}~\bibnamefont{Huysse}},
  \bibinfo{author}{\bibfnamefont{G.}~\bibnamefont{Reusen}},
  \bibinfo{author}{\bibfnamefont{P.}~\bibnamefont{Van~Duppen}},
  \bibnamefont{and} \bibinfo{author}{\bibfnamefont{J.}~\bibnamefont{Wauters}},
  \bibinfo{journal}{Phys. Lett. B} \textbf{\bibinfo{volume}{226}},
  \bibinfo{pages}{27} (\bibinfo{year}{1989}).

\bibitem[{\citenamefont{Dracoulis et~al.}(1999)\citenamefont{Dracoulis, Byrne,
  Baxter, Davidson, Kib\'edi, McGoram, Bark, and Mullins}}]{dracoulis99}
\bibinfo{author}{\bibfnamefont{G.~D.} \bibnamefont{Dracoulis}},
  \bibinfo{author}{\bibfnamefont{A.~P.} \bibnamefont{Byrne}},
  \bibinfo{author}{\bibfnamefont{A.~M.} \bibnamefont{Baxter}},
  \bibinfo{author}{\bibfnamefont{P.~M.} \bibnamefont{Davidson}},
  \bibinfo{author}{\bibfnamefont{T.}~\bibnamefont{Kib\'edi}},
  \bibinfo{author}{\bibfnamefont{T.~R.} \bibnamefont{McGoram}},
  \bibinfo{author}{\bibfnamefont{R.~A.} \bibnamefont{Bark}}, \bibnamefont{and}
  \bibinfo{author}{\bibfnamefont{S.~M.} \bibnamefont{Mullins}},
  \bibinfo{journal}{Phys. Rev. C} \textbf{\bibinfo{volume}{60}},
  \bibinfo{pages}{014303} (\bibinfo{year}{1999}).

\bibitem[{\citenamefont{Hellemans}(2007)}]{hellemans07b}
\bibinfo{author}{\bibfnamefont{V.}~\bibnamefont{Hellemans}}, Ph.D. thesis,
  \bibinfo{school}{Ghent University} (\bibinfo{year}{2007}).

\bibitem[{\citenamefont{Rowe}(1970)}]{rowe70}
\bibinfo{author}{\bibfnamefont{D.~J.} \bibnamefont{Rowe}},
  \emph{\bibinfo{title}{Nuclear Collective Motion, Models and Theory}}
  (\bibinfo{publisher}{Methuen}, \bibinfo{address}{London},
  \bibinfo{year}{1970}).

\bibitem[{\citenamefont{Dracoulis}(1994)}]{dracoulis94}
\bibinfo{author}{\bibfnamefont{G.~D.} \bibnamefont{Dracoulis}},
  \bibinfo{journal}{Phys. Rev. C} \textbf{\bibinfo{volume}{49}},
  \bibinfo{pages}{3324} (\bibinfo{year}{1994}).

\bibitem[{\citenamefont{Allatt}(1998)}]{allatt98}
\bibinfo{author}{\bibfnamefont{R.~G.} \bibnamefont{Allatt}},
  \bibinfo{journal}{Phys. Lett. B} \textbf{\bibinfo{volume}{437}},
  \bibinfo{pages}{29} (\bibinfo{year}{1998}).

\bibitem[{\citenamefont{Page et~al.}(2003)\citenamefont{Page, Ackerman,
  Andreyev, Cagarda, Eskola, Gerl, Greenless, Heissberger, Heyde, Hofmann
  et~al.}}]{page01}
\bibinfo{author}{\bibfnamefont{R.~D.} \bibnamefont{Page}},
  \bibinfo{author}{\bibfnamefont{D.}~\bibnamefont{Ackerman}},
  \bibinfo{author}{\bibfnamefont{A.~N.} \bibnamefont{Andreyev}},
  \bibinfo{author}{\bibfnamefont{P.}~\bibnamefont{Cagarda}},
  \bibinfo{author}{\bibfnamefont{K.}~\bibnamefont{Eskola}},
  \bibinfo{author}{\bibfnamefont{J.}~\bibnamefont{Gerl}},
  \bibinfo{author}{\bibfnamefont{P.~T.} \bibnamefont{Greenless}},
  \bibinfo{author}{\bibfnamefont{F.~P.} \bibnamefont{Hessberger}},
  \bibinfo{author}{\bibfnamefont{K.}~\bibnamefont{Heyde}},
  \bibinfo{author}{\bibfnamefont{S.}~\bibnamefont{Hofmann}}
  \bibnamefont{\textit{et~al.}}, in \emph{\bibinfo{booktitle}{Proceedings of the 3rd
  International Conference on Exotic Nuclei and Atomic Masses ENAM 2001}}, edited by
  \bibinfo{editor}{\bibfnamefont{J.}~\bibnamefont{\"Ayst\"o}},
  \bibinfo{editor}{\bibfnamefont{P.}~\bibnamefont{Dendooven}},
  \bibinfo{editor}{\bibfnamefont{A.}~\bibnamefont{Jokinen}}, \bibnamefont{and}
  \bibinfo{editor}{\bibfnamefont{M.}~\bibnamefont{Leino}}
  (\bibinfo{publisher}{Springer-Verlag, Berlin/Heidelberg},
  \bibinfo{year}{2003}), p. \bibinfo{pages}{309}.

\bibitem[{\citenamefont{Pakarinen}(2005)}]{pakarinen05b}
\bibinfo{author}{\bibfnamefont{J.}~\bibnamefont{Pakarinen}}, Ph.D. thesis,
  \bibinfo{school}{University of Jyv\"askyl\"a} (\bibinfo{year}{2005}).

\end{thebibliography}

\end{document}